\documentclass[sigconf]{acmart}
\makeatletter
\def\@ACM@checkaffil{
    \if@ACM@instpresent\else
    \ClassWarningNoLine{\@classname}{No institution present for an affiliation}%
    \fi
    \if@ACM@citypresent\else
    \ClassWarningNoLine{\@classname}{No city present for an affiliation}%
    \fi
    \if@ACM@countrypresent\else
        \ClassWarningNoLine{\@classname}{No country present for an affiliation}%
    \fi
}
\makeatother

\AtBeginDocument{%
  \providecommand\BibTeX{{%
    \normalfont B\kern-0.5em{\scshape i\kern-0.25em b}\kern-0.8em\TeX}}}

\setcopyright{acmcopyright}
\copyrightyear{2024}
\acmYear{2024}
\acmDOI{XXXXXXX.XXXXXXX}

\acmConference[]{}{}{}
\acmPrice{15.00}
\acmISBN{978-1-4503-XXXX-X/18/06}

\usepackage{booktabs}
\usepackage{multirow}
\usepackage{pifont}
\usepackage{titlesec}
\usepackage{enumitem}

\setcopyright{none}
\copyrightyear{}
\acmYear{}
\acmDOI{}
\renewcommand\footnotetextcopyrightpermission[1]{}

\begin{document}

\title{\includegraphics[height=\baselineskip]{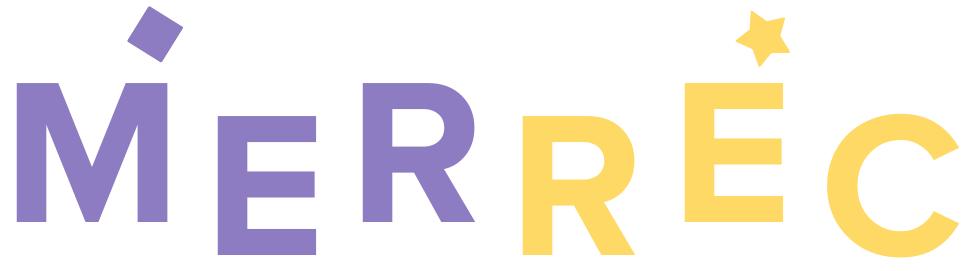}: A Large-scale Multipurpose Mercari Dataset for Consumer-to-Consumer Recommendation Systems}


\author{Lichi Li}
\authornote{Equal contribution}
\authornote{Work done while at Mercari.}
\affiliation{%
  \institution{Independent Researcher}
}
\email{lichili233@gmail.com}

\author{Zainul Abi Din}
\affiliation{%
  \institution{Independent Researcher}
}
\email{zain.da819@gmail.com}
\authornotemark[1]
\authornotemark[2]

\author{Zhen Tan}
\affiliation{%
  \institution{Arizona State University}
}
\email{ztan36@asu.edu}
\authornotemark[1]

\author{Sam London}
\affiliation{%
  \institution{Independent Researcher}
}
\email{Sl.cashew@gmail.com}
\authornotemark[2]

\author{Tianlong Chen}
\affiliation{%
  \institution{University of North Carolina at Chapel Hill, MIT, Harvard University}
}
\email{tianlong@mit.edu}

\author{Ajay Daptardar}
\affiliation{%
  \institution{Mercari, Inc.}
}
\email{ajay@mercari.com}

\renewcommand{\shortauthors}{Li and Din, et al.}

\begin{abstract}
In the evolving e-commerce field, recommendation systems crucially shape user experience and engagement. The rise of Consumer-to-Consumer (C2C) recommendation systems, noted for their flexibility and ease of access for customer vendors, marks a significant trend. However, the academic focus remains largely on Business-to-Consumer (B2C) models, leaving a gap filled by the limited C2C recommendation datasets that lack in item attributes, user diversity, and scale. The intricacy of C2C recommendation systems is further accentuated by the dual roles users assume as both sellers and buyers, introducing a spectrum of less uniform and varied inputs. Addressing this, we introduce \textit{MerRec}, the first large-scale dataset specifically for C2C recommendations, sourced from the Mercari e-commerce platform, covering millions of users and products over 6 months in 2023. \textit{MerRec} not only includes standard features such as \texttt{user\_id}, \texttt{item\_id}, and \texttt{session\_id}, but also unique elements like timestamped action types, product taxonomy, and textual product attributes, offering a comprehensive dataset for research. This dataset, extensively evaluated across four recommendation tasks, establishes a new benchmark for the development of advanced recommendation algorithms in real-world scenarios, bridging the gap between academia and industry and propelling the study of C2C recommendations. Our experiment code is available at \url{https://github.com/mercari/mercari-ml-merrec-pub-us} and dataset at \url{https://huggingface.co/datasets/mercari-us/merrec}.
\end{abstract}




\maketitle

\section{Introduction}
In the era of flourishing e-commerce, recommendation systems have emerged as pivotal tools in shaping user experience and driving customer engagement. These systems are instrumental in navigating the vast array of products and services offered online, making them a cornerstone of modern e-commerce platforms~\cite{isinkaye2015recommendation,zhao2021recbole,fan2022graph,zhang2019deep}. Currently, the literature of recommendation systems has been primarily dominated by Business-to-Consumer (B2C) models~\cite{alamdari2020systematic}, which have been extensively studied and refined over the years. These models typically revolve around businesses recommending professionally-sold products to general public individual consumers, leveraging structured data and well-defined user behavior patterns~\cite{li2015online}.

\begin{figure}[t]
  \centering
  \includegraphics[width=\linewidth]{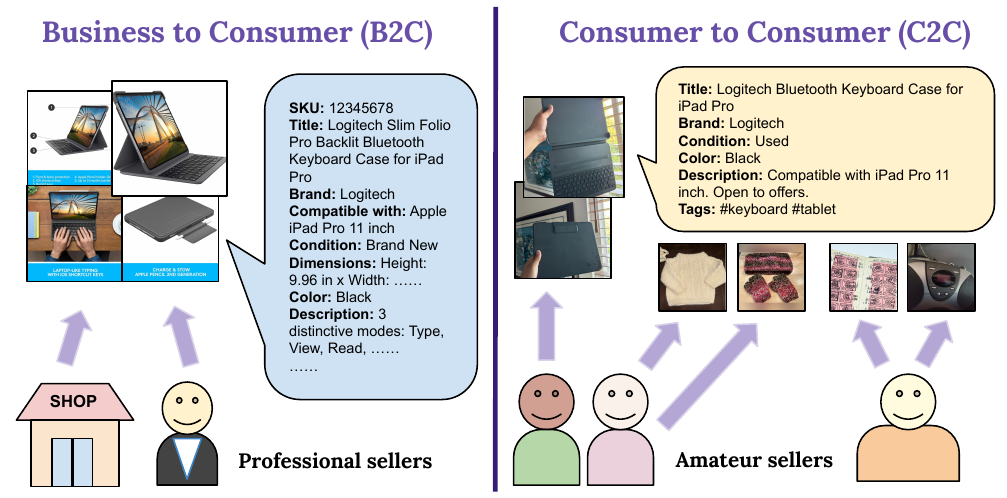}
  \caption{Comparison of B2C and C2C E-commerce Platforms: This illustration highlights the differences in product descriptions between B2C and C2C platforms. B2C platforms typically feature consistent and professionally crafted merchandise descriptions. Conversely, product descriptions on C2C platforms are often more varied and less standardized, posing challenges to the robustness of recommendations.}
  \label{fig:b2c_vs_c2c}
  \vspace{-10mm}
\end{figure}

In more recent years, with the advent of digital marketplaces and the shift towards more inclusive and flexible e-commerce platforms, the Consumer-to-Consumer (C2C)~\cite{ovaska2024green} business model has developed prosperously. In C2C e-commerce, individuals are both the sellers and buyers of products, creating a dynamic and multifaceted marketplace. This shift marks a significant trend for E-commerce, highlighting the need for recommendation systems that cater to these unique environments. Unlike their B2C counterparts, C2C platforms present unique challenges due to their unstructured nature, the dual roles of users, and the potential absence of conventional identifiers like Stock Keeping Units (SKUs). 

Yet, the academic community has not fully turned its attention to the intricacies of C2C models. This has left a noticeable gap between the real-world industry and academia, particularly in terms of datasets that adequately capture the complexities of C2C marketplace. The few existing C2C datasets that are available are often not designed with recommendation systems in mind~\cite{C2CFashion} or are hindered by limitations such as insufficient item detail, limited user diversity, and inadequate scale for the effective training and testing of sophisticated recommendation algorithms~\cite{ma2018entire,ebay2021,ovaska2024green}.

Addressing these challenges, our research introduces \textit{MerRec}, a groundbreaking dataset specifically tailored for C2C recommendation systems. Sourced from Mercari\footnote{\url{https://www.mercari.com/}},
a leading C2C e-commerce platform, MerRec stands out due to its \textbf{large-scale} nature, encompassing diverse consumer interactions, behaviors, and preferences within the Mercari marketplace. Moreover, the dataset, in addition to standard attributes like \texttt{user\_id}, \texttt{item\_id}, and \texttt{session\_id}, also incorporates \textbf{encompassing features} such as timestamped action types, detailed product taxonomy, and textual product attributes. These elements provide a rich tapestry of data, allowing for in-depth analysis of C2C marketplace dynamics.

Uniquely, MerRec is designed with the flexibility to accommodate the \textbf{fluid nature} of online C2C marketplaces, where item listings are continuously created, updated, and removed. This dynamic item identification supports the development of adaptive recommendation systems that remain effective in the ever-changing landscape of C2C e-commerce. With the dataset, we introduce a prototype model, \textit{Mercatran}, which demonstrates initial performance benchmarks for handling non-SKU item identifications within this novel context.

The comprehensiveness makes MerRec a multipurpose dataset and can be utilized for diverse tasks related to recommendations. Here we focus on four different tasks: \ding{182}~\textit{Click-through rate (CTR) prediction}, which is a classical task where a recommendation model learns to anticipate whether a user click is likely to happen given some context. \ding{183}~\textit{Session-based Recommendation (SBR)}, or sequential recommendation, which is a widely used modeling technique to predict the next item, given a list of interacted items. \ding{184}~\textit{Multi-task Learning for Recommendation (MLR)}, where the recommendation model is expected to predict different types of user actions such as \verb|item_view| and \verb|item_like|. \ding{185}~\textit{Inference Acceleration for Recommendation (IAR)}, which aims to train modules to provide speedup to existing session-based recommendation models. Our benchmarks include both sophisticated baseline models and straightforward heuristic approaches, providing a comprehensive overview of MerRec's utility as an analytical testbed.

In conclusion, MerRec marks a significant advancement in the study of C2C recommendation systems. Its extensive scale, focus on C2C interactions, and rich feature set make it an essential resource for both researchers and practitioners in the field. MerRec not only addresses the current limitations in C2C recommendation research but also opens new pathways for innovation, paving the way for the development of more nuanced and effective recommendation systems in the vibrant world of e-commerce.

\section{Related Work}

\subsection{E-commerce Recommendation Systems} As the core for e-commerce platforms, recommendation systems have been extensively explored, with a predominant focus on B2C models. These systems leverage algorithms such as collaborative filtering~\cite{schafer2007collaborative}, content-based filtering~\cite{thorat2015survey}, and hybrid methods~\cite{burke2002hybrid} to suggest products to users based on past interactions and preferences. Seminal works in this domain have laid the foundation for understanding user behavior and personalizing the shopping experience in online platforms. Notably, research has delved into various aspects of recommendation systems, including algorithmic efficiency, personalization accuracy, and scalability to accommodate large user bases and diverse item catalogs.

The emergence of Consumer-to-Consumer (C2C) platforms like eBay\footnote{\url{https://www.ebay.com/}}, Etsy\footnote{\url{https://www.etsy.com/}}, and Mercari has significantly altered the dynamics of the e-commerce ecosystem. These platforms empower individuals to both offer and acquire goods directly, fostering a marketplace characterized by its vast diversity yet complicated by intricate user interactions and varied transaction models. Research into C2C platforms has delved into the socio-economic behaviors of users, the mechanisms of trust and reputation systems, and the distinct challenges posed by a marketplace driven by its participants~\cite{cano2023sustainable}. Despite this, crafting recommendation systems for such environments is fraught with challenges. The lack of uniform product identifiers (SKUs), the dynamic nature of product listings, and the amalgamated roles of users as both vendors and consumers add layers of complexity. Crucially, the absence of a comprehensive, large-scale dataset presents a significant obstacle for researchers in the field, hindering the advancement of tailored recommendation solutions.

\subsection{Datasets for Recommendation Systems} The development of advanced recommendation systems critically hinges on the availability of robust, comprehensive datasets. In the realm of B2C (Business-to-Consumer) scenarios, several well-established datasets are frequently employed. Notable examples include the Amazon product review dataset [1, 2], the Netflix Prize dataset [3], and the MovieLens dataset [4], which have all been foundational in advancing B2C recommendation system research.

In contrast, the landscape for C2C (Consumer-to-Consumer) recommendation systems is notably underrepresented in available research resources. The specific challenges of C2C environments—such as more diverse transaction types and highly individualized user interactions—are not adequately captured by traditional B2C datasets. While some researchers have attempted to adapt B2C datasets to model C2C scenarios, these adaptations often fail to fully replicate the unique dynamics of actual C2C marketplaces [5, 6]. Moreover, datasets explicitly designed for C2C recommendation systems are scarce. The few that exist, such as those derived from smaller, niche online marketplaces [7], tend to offer only a limited view of the markets, lacking the scale and diversity seen in mainstream C2C platforms [8].

This gap in dataset availability underscores a critical need in the field: the development of large-scale, diverse datasets that are tailor-made for C2C recommendation systems. Such datasets would not only enhance the realism and applicability of research but also pave the way for innovations that are finely attuned to the nuances of consumer-driven marketplaces [9].

\begin{table}[]
\caption{Statistics and Explanation of all included features in MerRec. ``N/A''s indicates the exact counts are not calculated since they are less meaningful. Counts for \texttt{event\_id} and all features marked with ``tokens'' are non-distinct occurrences.}
\label{tab:stats_explanation}
\resizebox{\linewidth}{!}{%
\begin{tabular}{@{}ccc@{}}
\toprule
\textbf{Feature}               & \textbf{Distinct Counts}& \textbf{Description}                                            \\ \midrule
\texttt{user\_id} &
  5,569,367 &
  Globally unique user account ID. \\
\texttt{sequence\_id} &
  69,144,727 &
  User-level unique sequence ID. \\
\texttt{session\_id} &
  227,167,616 &
  User-level unique session ID. \\
\texttt{stime}                 &                          N/A& Timestamp in UTC timezone.                                      \\
\texttt{sequence\_length}      &                          N/A& Number of events within this \texttt{sequence\_id}.             \\
\texttt{event\_id}             & 1,274,814,848            & Action event ID.\\
\texttt{item\_id}              & 83,078,407               & Globally unique item ID, representing a single listing post.   \\
\texttt{product\_id}           & 1,403,098                & A concatenation between \texttt{brand\_id} and \texttt{c2\_id}. \\
\texttt{name}                  & 8,203,775,179 tokens     & Title text of the item.                                         \\
\texttt{price}                 &                          N/A& Price of the item represented in USD.                           \\
\texttt{c0\_name}              & 1,925,154,049 tokens     & Text label of the item's c0-level category.                     \\
\texttt{c0\_id}                & 17                       & ID of the item's c0-level category.                             \\
\texttt{c1\_name}              & 2,679,629,082 tokens     & Text label of the item's c1-level category.                     \\
\texttt{c1\_id}                & 309                      & ID of the item's c1-level category.                             \\
\texttt{c2\_name}              & 1,989,892,371 tokens     & Text label of the item's c2-level category.                     \\
\texttt{c2\_id}                & 3073                     & ID of the item's c2-level category.                             \\
\texttt{brand\_name}           & 1,554,523,806 tokens     & Text label of the item's brand.                                 \\
\texttt{brand\_id}             & 20,001                   & ID of the item's brand.                                         \\
\texttt{item\_condition\_name} & 1,595,094,363 tokens     & Text label of the item's condition (e.g. New, Good, Fair).      \\
\texttt{item\_condition\_id}   & 6                        & ID of the item's condition.                                     \\
\texttt{size\_name}            & 916,484,536 tokens       & Text label of the item's size (i.e. for shoes, jackets, pants). \\
\texttt{size\_id}              & 355                      & ID of the item's size.                                          \\
\texttt{shipper\_name}         & 2                        & Binary text label, \textit{i.e.}, Buyer or Seller.\\
\texttt{shipper\_id}           & 2                        & Binary ID of the item's shipment-paying party.                  \\ \bottomrule
\end{tabular}%
}
\end{table}

\begin{figure}[b]
  \centering
  \includegraphics[width=\linewidth]{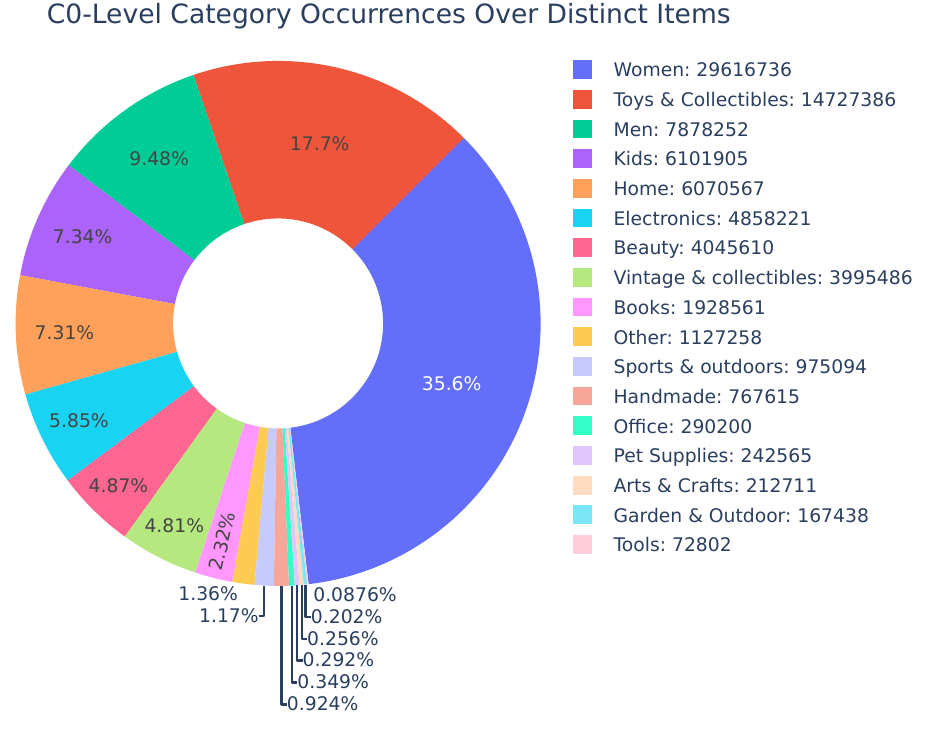}
  \caption{Break down of the C0-level category appearances over the distinct items. MerRec dataset has some concentration over Women and Toys \& Collectibles, but is overall reasonably balanced to represent a broad spectrum of items available on Mercari marketplace.}
  \label{fig:c0_counts}
  \vspace{-2mm}
\end{figure}

\section{The \includegraphics[height=\baselineskip]{figures/logo.pdf} Dataset}
\label{data_description}

Mercari, an online Consumer-to-Consumer (C2C) marketplace, facilitates a unique environment where users can seamlessly transition between the roles of buyer and seller, minimizing logistical hurdles. This research is primarily focused on analyzing buyer-side preferences to enhance the relevancy of recommendation tasks within this platform.

From a seller's perspective, Mercari requires the provision of detailed information for each listing, which may include individual items or bundles. Essential details such as titles, brands, categories, images, shipping cost responsibilities, price, and condition are mandated, while size and color are optional. These details are dynamic, allowing sellers to update their listings, which introduces variability in how items are represented over time.

The nature of Mercari's seller base, predominantly composed of non-professional sellers, introduces unique challenges not typically encountered on Business-to-Consumer (B2C) platforms. The primary issue is the variability in the quality and consistency of the listings due to the self-reported nature of the information. This can lead to inaccuracies in brand or category labels, incomplete or generic item descriptions, and sometimes, the omission of crucial details like size or color. Additionally, the absence of standardized identifiers like Stock Keeping Units (SKUs) complicates item identification, as each item on Mercari is unique and cannot be repurchased once sold, unlike traditional B2C platforms where items can be sold multiple times.

On the buyer side, Mercari offers a diverse range of user interface (UI) options for item discovery, with users able to interact with items through various actions such as clicking, liking, adding to cart, making offers, initiating, and completing transactions. These interactions serve as indicators of user interest and intent, forming the basis for our analysis in the MerRec dataset.

The MerRec dataset is meticulously designed to capture the intricate and evolving dynamics of user behavior and item characteristics on Mercari. It aims to establish a comprehensive understanding of the relationships between users and items by analyzing user actions in conjunction with item details. This approach seeks to address the challenges posed by the C2C business model, particularly the variability of item descriptions and the absence of standard identifiers, to enhance the effectiveness of recommendation systems in such a unique marketplace environment.

\begin{figure*}[t]
  \centering
  \includegraphics[width=\linewidth]{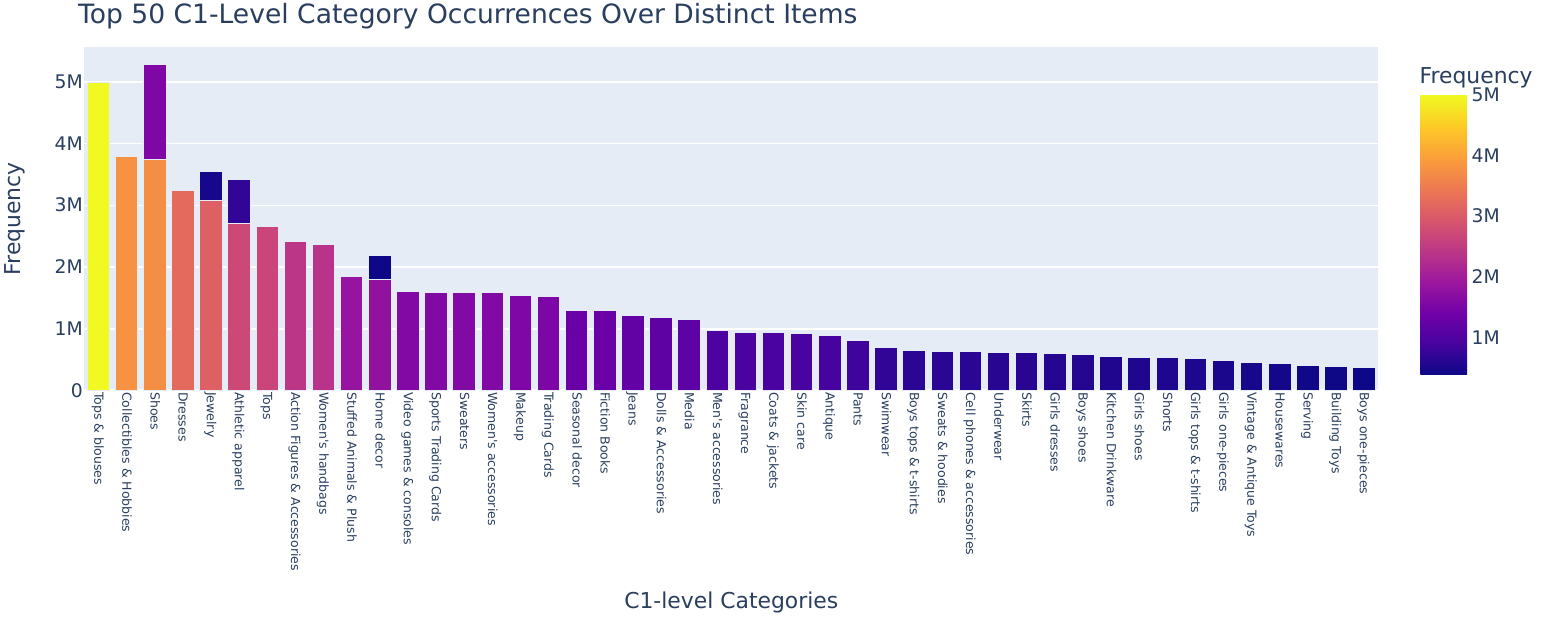}
  \caption{Break down of the top 50 C1-level category appearances over the distinct items. The stacked bars represent categories which have the same name but originally belonging under different C0-level categories. For example, there are two distinct C1 category IDs called \textit{Shoes}, one from the C0 \textit{Womens} and another from the C0 \textit{Mens}.}
  \label{fig:c1_top50_counts}
  \vspace{-1mm}
\end{figure*}

\begin{figure}
  \centering
  \includegraphics[width=\linewidth]{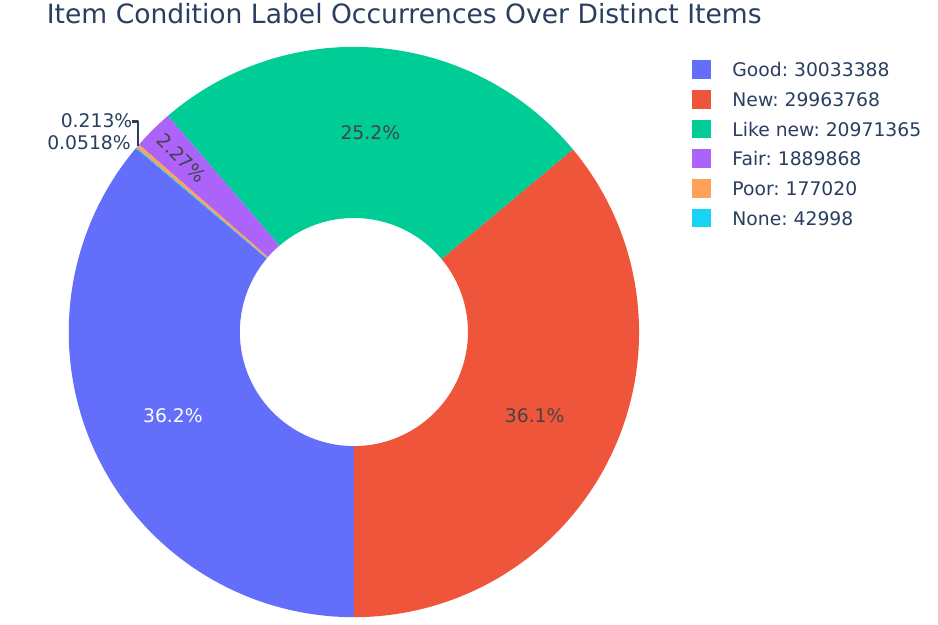}
  \caption{Break down of item condition appearances over the distinct items.}
  \label{fig:condition_counts}
\end{figure}

\subsection{Dataset Requirement}

To ensure the MerRec dataset is a valuable resource for analyzing complex and diverse user interactions with content on Mercari, we established several criteria for its composition. These criteria are designed to encapsulate a wide spectrum of user activities and item specifics, enhancing the dataset's utility for research purposes:

\begin{itemize}[leftmargin=*]
\item \textbf{Item Variety:} The dataset should encompass a wide array of item types to represent the vast diversity of products available on the Mercari platform. This variety allows for a comprehensive analysis of user preferences and behaviors across categories.

\item \textbf{User Actions:} It's essential to capture a broad range of user actions to understand how users interact with the platform. This includes not just the final purchase but also preliminary actions like browsing, liking, adding items to carts, and making offers, which collectively offer insights into the user's journey.

\item \textbf{Item Details:} Detailed information about each item is crucial. This encompasses not just basic descriptors like title, category, and price but also more nuanced details such as condition, size, and color. Such granularity enables a deeper understanding of what influences user preferences and decisions.

\item \textbf{Contextual Information:} Each user action should be accompanied by contextual data, such as the time of the action and the type of action performed. This temporal and qualitative context adds a layer of depth to the analysis, allowing for temporal trends and patterns in user behavior to be identified.

\item \textbf{Recency:} The dataset should prioritize recent items and user actions to ensure the data reflects current trends and preferences on the platform. This recency is vital for developing recommendation systems that are aligned with the latest user interests and marketplace dynamics.

\end{itemize}

\begin{figure}[t]
\vspace{-2mm}
  \centering
  \includegraphics[width=\linewidth]{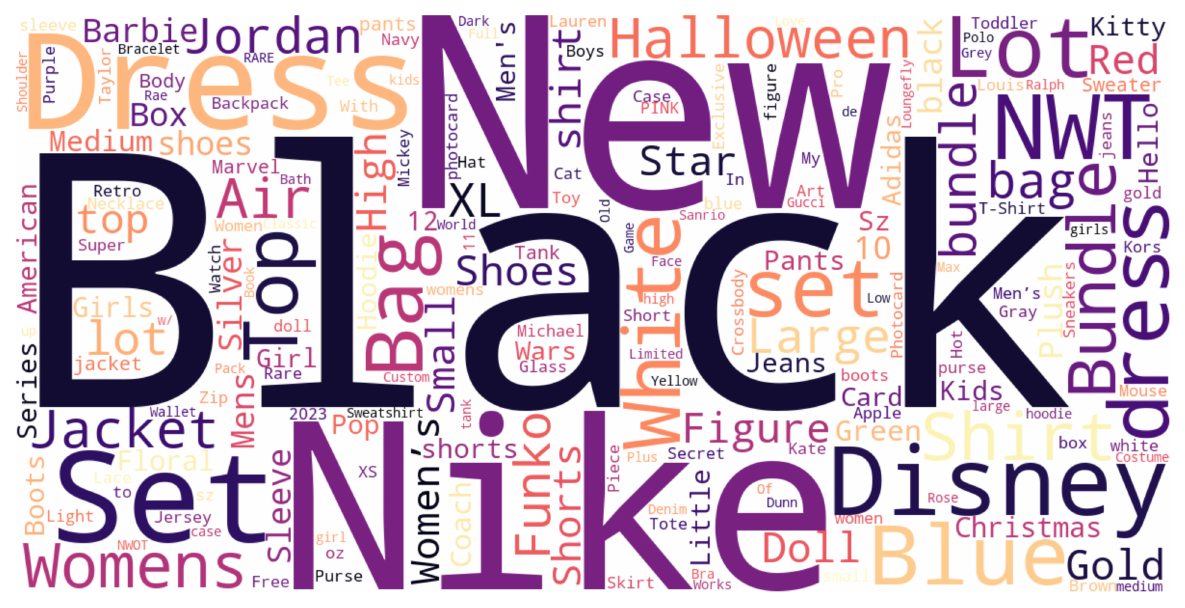}
  \caption{Word cloud of the most frequently observed words within MerRec's item titles. Stop words are omitted.}
  \label{fig:title_wordcloud}
  \vspace{-8mm}
\end{figure}

\noindent Furthermore, it is imperative that the dataset adheres to all applicable legal and privacy regulations to ensure ethical usage and respect for user confidentiality. By meeting these requirements, the MerRec dataset aims to provide a robust foundation for academic and practical research into recommendation systems within the dynamic and user-driven context of C2C e-commerce platforms that are similar to Mercari. The basic statistics and explanations for included features are included in Table~\ref{tab:stats_explanation}.

To ensure comprehensive event representation, our dataset does not limit the origins of user interactions within the Mercari platform's user interface, encompassing both mobile and web clients. This means an interaction, such as a click, could originate from various sections like the home page's ``For You'' or ``Recently Viewed'', 
or within a search result browsing session, among others. Similarly, actions like adding an item to a cart, initiating a checkout, or liking an item can occur across multiple interface points. These interactions are influenced by a diverse array of underlying recommendation, retrieval, and ranking systems, contributing to the varied contexts captured in our data. To capture a diverse range of user activities over time, our dataset includes sampled user events from different months, adhering to size limits per month to manage the volume of sequences.

Our dataset does not explicitly filter item types, except certain categories that were deprecated due to changes in Mercari's business practices. This approach ensures the preservation of item diversity, reflecting the initial sampling scope. Mercari defines its item category in a 3-level tree structure, ranging from c0 (least granular level), c1 (intermediate level) to c2 (most granular level), and this dataset showcases example items from most of them. See Figure~\ref{fig:c0_counts}, \ref{fig:c1_top50_counts}, \ref{fig:condition_counts}, and \ref{fig:title_wordcloud} for illustration. More are presented in the Appendix.

We intentionally designed the MerRec dataset to encapsulate the intricate dynamics of user behavior across multiple sessions, rather than confining analysis to isolated user interactions within single sessions. This approach is grounded in the nuanced patterns observed in Mercari's live environment, where user preferences and interests can fluctuate significantly over time. Users might pivot to entirely different product categories in subsequent sessions or persistently explore similar items, indicating both transient and enduring interests. To capture this breadth of user engagement, each user action is meticulously tagged with a session ID, enabling a layered analysis that discerns patterns within discrete sessions while tracing the evolution of user interests across sessions. This multifaceted tracking fosters a deeper understanding of user behavior, revealing how interests either diversify or stabilize over time, and provides invaluable insights into the dynamic interplay between user preferences and marketplace offerings on Mercari.

\subsection{Data Cleaning and Processing}
\label{data_description:cleaning}

\begin{figure}
  \centering
  \includegraphics[width=\linewidth]{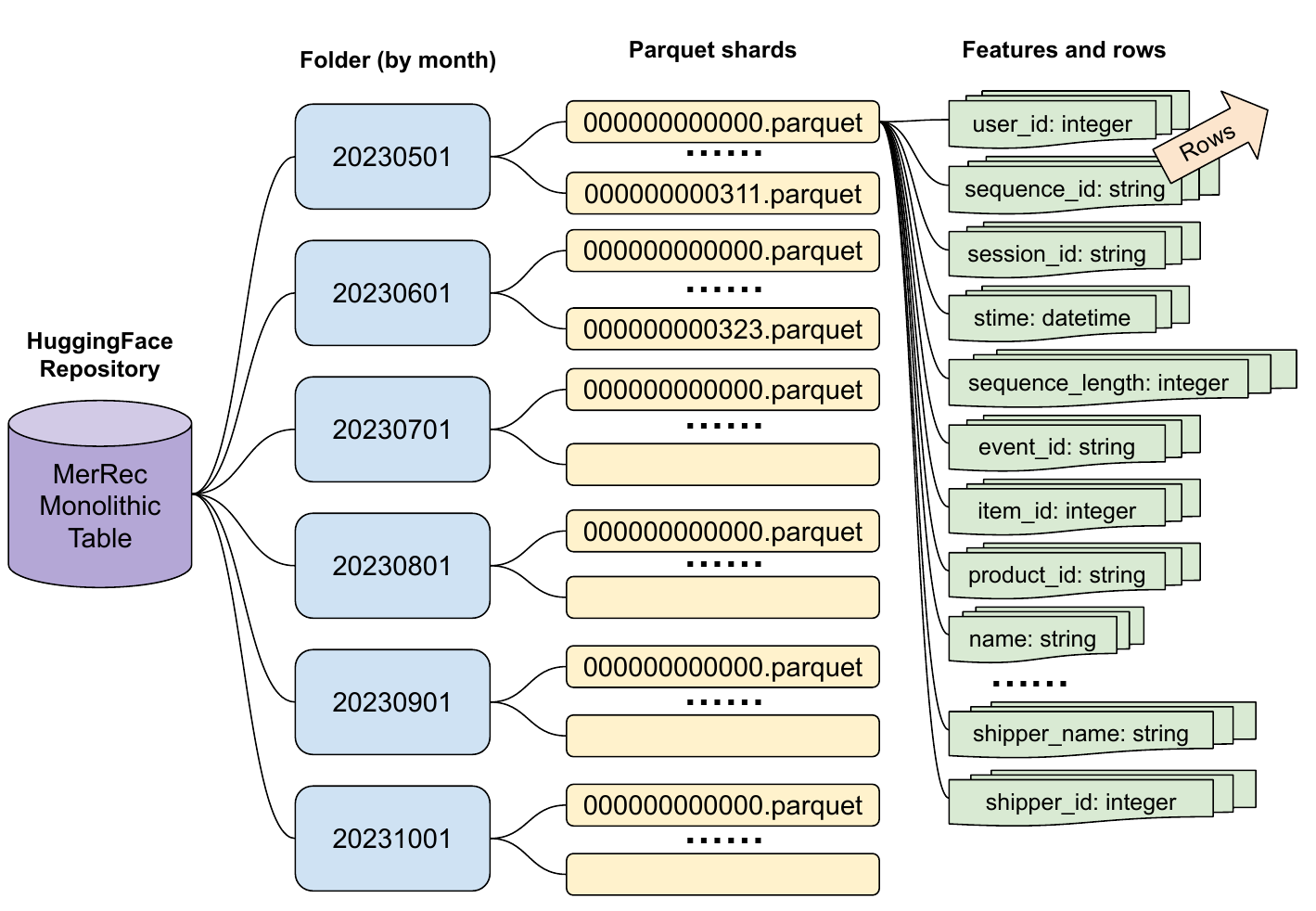}
  \caption{Illustrative overview of the dataset organization.}
  \label{fig:dataset_structure}
\vspace{-0.5cm}
\end{figure}

To enhance the quality of the MerRec dataset, a meticulous data cleaning and processing protocol was implemented. This involved the following key steps:
\vspace{-0.1cm}
\begin{enumerate}[leftmargin=*]
\item \textbf{User and Item Filtering:} Initial cleaning focused on removing users with banned or terminated accounts and items that breached platform service rules, such as those containing illegal content or crafted information to exploit the system with misleading information.

\item \textbf{Sequence Segmentation:} Given the observed long-tail distribution in user sequence lengths, longer sequences were segmented into shorter, fixed-length segments to standardize the data structure. This segmentation facilitates analysis, though researchers are encouraged to reconstruct the original sequences if required for specific studies.

\item \textbf{Redundancy Reduction:} To address the issue of repetitive actions within sequences, such as consecutive clicks on the same item, a deduplication process was applied. This step ensured that only unique consecutive interactions were retained, reducing redundancy and enhancing data conciseness.

\item \textbf{Privacy Protection:} Rigorous measures were taken to safeguard user privacy. This included excluding users from certain regions to comply with local regulations, anonymizing all ID fields with pseudonymous identifiers, and converting all timestamps to a uniform UTC format to obscure original local time.

\item \textbf{SKU Alternative Exploration:} An innovative approach was adopted to explore alternatives to traditional Stock Keeping Units (SKUs) in the C2C context. A synthetic field named as \texttt{product\_id} was created by merging brand and the most detailed category ID available, offering a novel way to study item identification in the absence of SKUs. However, it's important to acknowledge the limitations of this approach, as it may not provide the granularity required for effective recommendation in a live marketplace setting. See Figure \ref{fig:iphone_prices} for an illustrative example of this constraint.

\begin{figure}
  \centering
  \includegraphics[width=\linewidth]{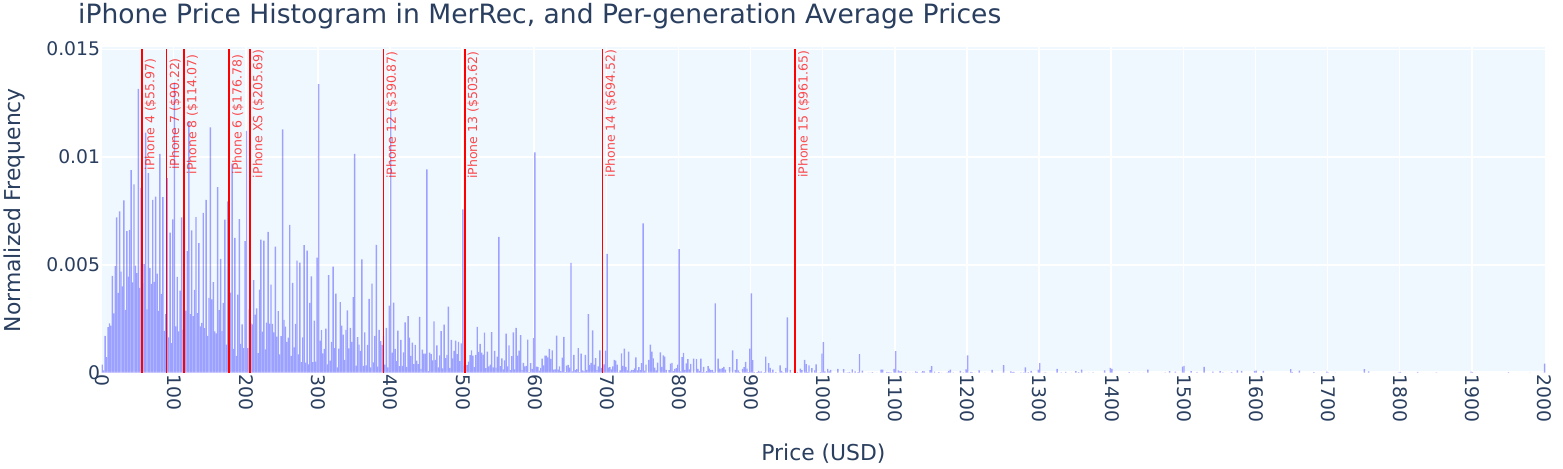}
  \caption{Average prices of different iPhone series (red) among all items (purple) within the same brand-C2 combination (i.e. \texttt{product\_id}) of Apple-Smartphones. Without SKU-tagging, this challenges the recommendation system to identify, differentiate and precisely promote the correct product. The price and product feature variation is high even in this example, and systems should not assume users to be broadly interested in all products under the same \texttt{product\_id}.}
  \label{fig:iphone_prices}
\vspace{-0.5cm}
\end{figure}

\end{enumerate}
The dataset is a single, monolithic table that is sharded into 6 folders, each folder corresponding to a specific month in 2023, where there are a few hundreds of parquet shards within each such folder. When loading and simply concatenating all parquet files inside one of the 6 folders, you have all the rows MerRec includes for that month. When concatenating all parquet files across all 6 folders, you then have the entire dataset\footnote{While we acknowledge that this format does imply some suboptimality in terms of raw disk storage size, we ultimately chose this format acknowledging the fact that (1) this dataset is practically large enough that sharding and batched loading is generally inevitable on broadly available training hardware or setup, and (2) each row is already pre-formatted in a way that conveniently supplies all the features in a flattened way, as well as (3) the items, when showing up multiple times throughout the dataset, may look different between the instances due to our dataset being a snapshot-oriented design (meaning items are faithfully represented back in time per each instance).}. Each single parquet file contains individual user actions, with each row detailing a unique interaction instance. Actions within a user's sequence can be reconstructed by aggregating data based on user and sequence identifiers and ordered by timestamp. See Figure \ref{fig:dataset_structure} for a structural overview.

This structured approach to data cleaning and processing ensures the MerRec dataset is a high-quality resource for studying user interactions on the Mercari platform, while also addressing privacy concerns and exploring innovative solutions to challenges inherent in C2C e-commerce environments.

\subsection{Comparison to Other E-commerce Datasets}

We compare MerRec against other open large-scale e-commerce recommendation datasets in Table \ref{tab:dataset_comparison}. To our best awareness, Retailrocket~\cite{zykov2022retailrocket} and MerRec are the only datasets in this list where item instances are back-in-time snapshots. Amongst the datasets, MerRec provides the highest number of unique items (+72\% against Amazon~\cite{hou2024amazonreviews}), interaction instances (+122\% against Amazon~\cite{hou2024amazonreviews}), number of unique interaction types (+50\% against DIGINETICA\footnote{\url{https://competitions.codalab.org/competitions/11161}}), most amount of categories (+103.6\% against Retailrocket~\cite{zykov2022retailrocket}), number of sessions (+2358\% against YOOCHOOSE\footnote{\url{https://recsys.acm.org/recsys15/challenge/}}), most generally feature-comprehensive by far while being the most up-to-date in recency. MerRec is also the most transparent in dataset production process among these by far.

\begin{table*}[]
\centering
\caption{Comparison with other large scale e-commerce datasets. Entries with explicit quoted numbers are either from claims of official work, or from our own calculation where feasible. Cells with question mark represent missing public information absent from original work or practically infeasible to gather or recover. Some columns (e.g. condition, shipment paying party) are omitted here to constrain the table size.}
\vspace{-1mm}
\label{tab:dataset_comparison}
\resizebox{\linewidth}{!}{%
\begin{tabular}{cccccccccccccccc}
\hline
\textbf{Dataset} &
  \textbf{Market Type} &
  \textbf{Users} &
  \textbf{Items} &
  \textbf{Interactions} &
  \textbf{No. Interaction Types} &
  \textbf{Categories / Leveled} &
  \textbf{Brand} &
  \textbf{Price} &
  \textbf{Color} &
  \textbf{Size} &
  \textbf{Timestamp} &
  \textbf{Item Tokens} &
  \textbf{SKU / UPC} &
  \textbf{Covered Year} &
  \textbf{Sessions} \\ \hline
MerRec (Ours)    & C2C (General) & 5.56M   & 83.07M  & 1.27B   & 6 & 3399 / Yes & Yes & Yes & Yes & Yes & Yes & 18.86B  & No  & Half of 2023 & 227.16M \\
Amazon \cite{hou2024amazonreviews}           & B2C (General) & 54.51M  & 48.19M  & 571.54M & 2 & 33 / No    & Yes & Yes & Yes & Yes & Yes & 30.78B  & Yes & 1996-2023    & No      \\
Tmall \cite{tmall2016}           & B2C (General) & 645.37K & 2.35M   & 44.52M  & 2 & 72 / No    & No  & No  & No  & No  & Yes & No      & Yes & Half of 2015 & 200.28K \\
Amazon-M2 \cite{jin2023amazon}        & B2C (General) & No      & 1.41M   & 16.79M  & 1 & No / No    & Yes & No  & Yes & Yes & No  & Yes     & Yes & ?            & 3.96M   \\
DIGINETICA       & ?             & 232.93K & 184.04K & 3.3M    & 4 & 1217 / No  & No  & Yes & No  & No  & Yes & 941.64K & ?   & ?            & 573.93K \\
YOOCHOOSE       & ?             & No      & 52.73K  & 34.15M  & 1 & 348 / No   & No  & Yes & No  & No  & Yes & No      & ?   & ?            & 9.24M   \\
Retailrocket \cite{zykov2022retailrocket}    & ?             & 1.4M    & 417.05K & 2.75M   & 3 & 1669 / Yes & No  & No  & No  & No  & Yes & 51.29M  & ?   & ?            & No      \\
Ali-CCP \cite{ma2018entire}        & C2C (General) & 400K    & 4.3M    & 87.41M  & 3 & ? / ?      & Yes & No  & No  & No  & No  & No      & ?   & ?            & No      \\
Alibaba-iFashion \cite{chen2019alibabaifashion} & C2C (Fashion) & 3.56M   & 4.46M   & 191.39M & 1 & 75 / No    & No  & No  & No  & No  & No  & 7.7M    & ?   & ?            & No      \\ \hline
\end{tabular}%
}
\end{table*}

\section{Experiment \& Analysis} \label{experiment}

In this section, we present the application of various machine learning and recommendation models to specific tasks using the MerRec dataset, demonstrating their effectiveness and performance outcomes. We refer to a previous work~\cite{tenrec2022} for the implementations of most of these models, with necessary modifications to accommodate the unique characteristics of the MerRec dataset.


\begin{figure}
  \centering
  \includegraphics[width=\linewidth]{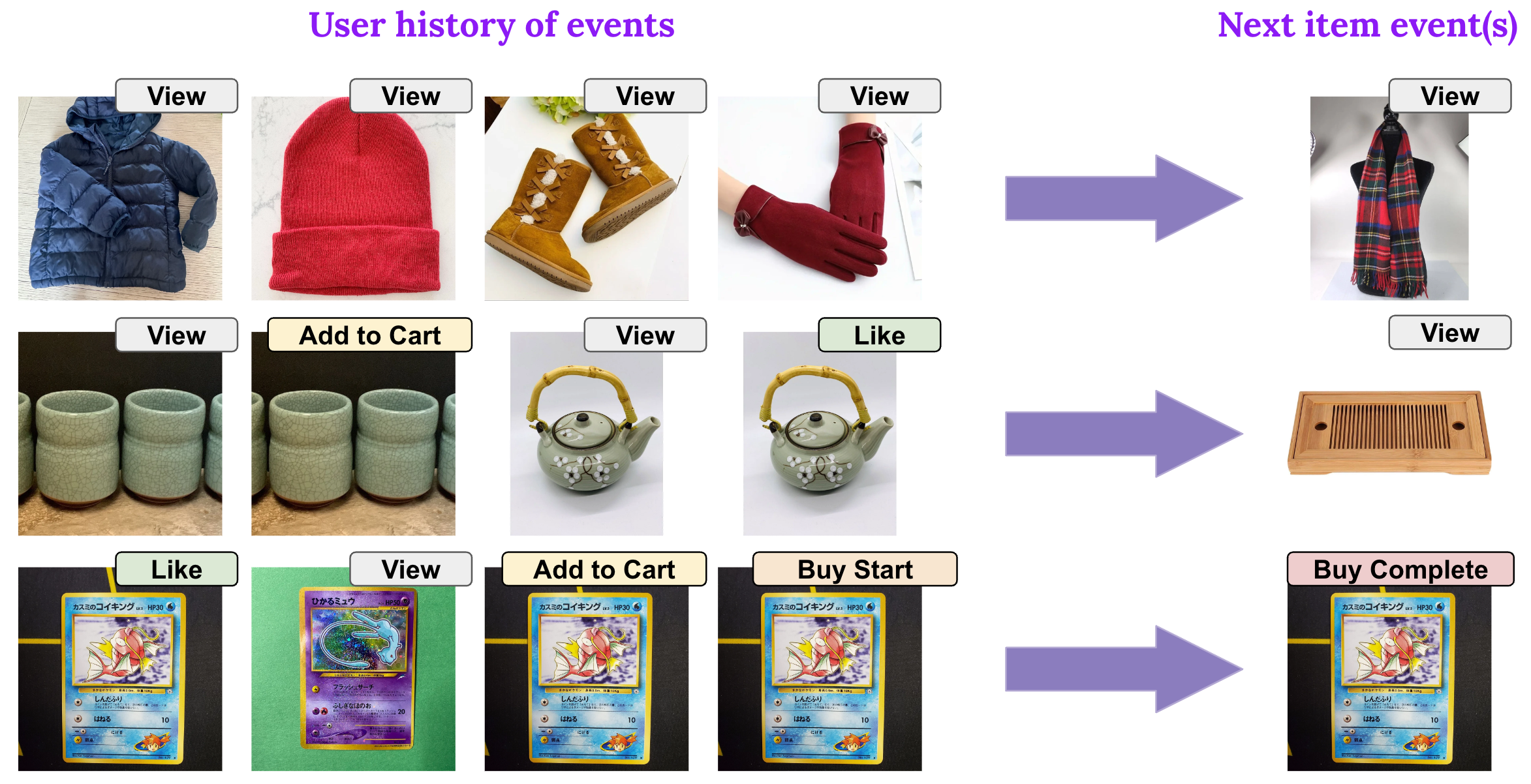}
  \caption{Illustrative examples of item action event sequences in MerRec. The separation between "history" or "next" is not explicitly enforced by the dataset but by the experiment tasks defined in Section \ref{experiment}.}
  \label{fig:event_sequence_flow_diagram}
\vspace{-0.5cm}
\end{figure}

\subsection{Click-Through Rate (CTR) Prediction} \label{experiment:ctr}

\subsubsection{Task Description.} CTR prediction is a fundamental task in recommendation systems, aiming to forecast the likelihood of a user clicking on an item. Utilizing the MerRec dataset, we explored this task by employing a range of competitive models, reformatted to leverage the rich user action and item attribute data available in MerRec. The models were tasked with predicting the occurrence of an \verb|item_view| action based on historical user interactions and item metadata, employing a binary classification approach.

\subsubsection{Dataset Setup.} The data was restructured into snapshots using a rolling window technique, allowing each model to make predictions based on a predefined historical context. Distinctively, our experiments did not restrict the types of user actions considered as input, diverging from some traditional CTR prediction approaches that focus exclusively on click events. Additionally, as MerRec does not include direct user demographic features (e.g. age, gender, ethnicity), we instead take advantage of the rich item features available to captivate dynamic interest nuance. This decision was based on the premise that various types of user actions can provide valuable signals about user preferences which may change at any moment throughout their platform activity, in ways less relatable to what their demographic background would have suggested. The dataset for CTR prediction, along with the multi-task learning (MTL) task, comprises a substantial volume of unique items, users, sequences, sessions, and product IDs.

To limit the total amount of computation time and resource, we chose to omit comprehensive hyperparameter search and reduce the sample used specifically for the benchmark executions. We use the first among six months of data in MerRec. We set the shortest input history window sequence in this execution to be 7, where we generate the binary prediction on the 8th item. Sequences equal to 8 events become one snapshot row and those longer than 8 events become multiple snapshot rows using the rolling window\footnote{We chose this format in line with common practices in the CTR literature, e.g. \cite{shen2017deepctr, tenrec2022}.}. This implies we did not pad sequences shorter than 8 events (which does not exist in MerRec) to become longer sequences unlike TenRec~\cite{tenrec2022}. Due to time constraints we did not specifically study the positive to negative sampling ratio on MerRec and how it affects training quality of the benchmark models, thus no negative sampling is enforced in this work and is left to the reader to explore.

\begin{table}[]
\caption{Results of Click-through-rate (CTR) prediction.}
\vspace{-0.3cm}
\label{tab:ctr}
\resizebox{\linewidth}{!}{%
\begin{tabular}{@{}ccccc@{}}
\toprule
\textbf{Model} & \textbf{AUC} & \textbf{LogLoss} & \textbf{Train+Val Time (Hrs)} & \textbf{VRAM (GB)} \\ \midrule
AFM     & 0.703  & 0.4114 & 21.6 & 5.12  \\
NFM     & 0.6498 & 0.5394 & 21.2 & 4.05  \\
DCN     & 0.6399 & 0.5657 & 20.8 & 4.06  \\
DCNv2   & 0.6209 & 0.6978 & 22.8 & 4.07  \\
DeepFM  & 0.642  & 0.5554 & 20.3 & 4.06  \\
xDeepFM & 0.6066 & 0.78   & 52.8 & 13.91 \\
W\&D    & 0.6626 & 0.4843 & 21.1 & 4.06  \\ \bottomrule
\end{tabular}%
}
\vspace{-0.2cm}
\end{table}

Furthermore, unlike some CTR benchmark demonstrations such as TenRec~\cite{tenrec2022}, we do not restrict the input window of events to only be clicks (in MerRec's case, not restricted to \verb|item_view|). This is because we believe there are valid content interest signals coming from other action events, and they may offer hints to subsequent actions. Due to the setup similarity, this CTR dataset is also reused in the multi-task learning (MTL) subsection \ref{experiment:mtl} as well. As a result, there are 30,221,983 unique items, 2,767,956 unique users, 9,809,155 sequences and 915,453 unique product IDs in this CTR demonstrative subset. We adopted an approximately 8:1:1 split ratio between training, validation and test sets, up to small round-offs. In this context, \verb|product_id| was preferred over \verb|item_id| for practical reasons related to model scalability and memory constraints. Despite this compromise, the chosen approach facilitated the completion of the experiments without encountering out-of-memory issues.

\noindent\subsubsection{Baseline Setup.} As there is a large literature history on CTR prediction, we only chose a few example models here such as Attention FM (AFM)~\cite{afm2017} which adopted attention-based pooling for factorization machines~\cite{fm2010}, Wide \& Deep~\cite{widedeep2016} which jointly trains wide and deep networks for modeling low and high order interactions differently, NeuralFM~\cite{nfm2017} which stacks neural networks on top of parameter-free feature interaction pooling, DCN~\cite{dcn2017} which introduced a cross network with varied degrees of polynomial approximations, DCNv2~\cite{dcnv22021} which brings low rank efficiency from mixture of experts~\cite{moe1991} to DCN, DeepFM~\cite{deepfm2017} which improves beyond Wide \& Deep with shared embeddings, as well as xDeepFM~\cite{deepfm2018} which improves modeling on high order explicit and implicit feature interactions, all of which were shown as strong baselines~\cite{openctrbench2021, tenrec2022, dcnv22021, fmfm2021}. Additional choices such as InterHAt~\cite{interhat2020}, AFN+~\cite{afn2020}, and LorentzFM~\cite{lorentzfm2020} exist but were omitted in this work as the additional performance gains were shown to be limited~\cite{tenrec2022, openctrbench2021}. Other works such as DUMN \cite{dumn2021}, DFN~\cite{dfn2021}, DCIN~\cite{dcin2023} were excluded due to their feature incompatibility with MerRec. Some models like DIN \cite{din2018}, DIEN~\cite{dien2019}, SDIM~\cite{sdim2022}, FiBiNet++~\cite{fibinetpp2023}, TIN~\cite{tin2023}, EulerNet~\cite{eulernet2023}, DualMLP~\cite{finalmlp2023} and FinalMLP~\cite{finalmlp2023} were omitted here merely due to time constraints. All executions in this subsection were ran on an Nvidia T4 GPU equipped linux system with 8 cores and 104 GB RAM on Google Cloud Platform\footnote{\url{https://cloud.google.com/}}. See Table \ref{tab:hyperparameter_ctr} for more details about the hyperparameters. See Table \ref{tab:ctr} for the list of performance scores on the test set. 

\noindent\subsubsection{Implications.} In our test, we can see that CTR is a challenging task on MerRec and many method approaches a similar band of performance under limited hyperparameter tuning. We find that a classic model Attention FM~\cite{afm2017} leads ahead other even more recent baselines. It also shows that models with cross networks may endure a lower performance or are less straightforward to tune for capturing the varied degrees of interactions in MerRec. This section underscores the adaptability of existing models to the nuanced and dynamic data presented by the MerRec dataset, offering insights into the potential for future research and development in recommendation systems within C2C marketplaces.

\subsection{Session-based Recommendation (SBR)} 
\label{experiment:sbr}

\subsubsection{Task Description.} In this subsection of our study, we delve into the task of sequential recommendation, aiming to predict a user's next item interaction based on their previous actions. This involves analyzing a sequence of actions, $S^u = (S^u_1, S^u_2, ..., S^u_{n-1})$, to forecast the forthcoming $n^{th}$ interaction, $S^u_n$. 

For the SBR task with MerRec data, we adapted the data to align with common practices in sequential recommendation models. Specifically, we utilized \verb|product_id| to represent item interactions and limited each user interaction sequence to the last 20 actions, applying front-padding for shorter sequences and excluding those with fewer than 10 interactions.

\begin{table}[t]
\caption{Results of SBR with MerRec data.}
\vspace{-0.2cm}
\label{tab:sbr_bench}
\resizebox{\columnwidth}{!}{%
\begin{tabular}{@{}ccccccc@{}}
\toprule
\textbf{Model} & \textbf{nDCG @5} & \textbf{nDCG @20} & \textbf{Recall @5} & \textbf{Recall @20} & \textbf{Tr./Val. Time (Hrs)} & \textbf{VRAM (GB)} \\ \midrule
NextItNet & 0.257 & 0.298 & 0.348 & 0.490 & 37.26 & 16.89 \\
Bert4Rec  & 0.142 & 0.176 & 0.199 & 0.318 & 40.01 & 17.01 \\
GRU4Rec   & 0.244 & 0.286 & 0.333 & 0.477 & 30.85 & 11.92 \\
SASRec    & 0.175 & 0.206 & 0.235 & 0.344 & 29.74 & 11.93 \\ \bottomrule
\end{tabular}%
}
\vspace{-0.5cm}
\end{table}

\subsubsection{Dataset Setup.} Similar to the experimental setup in Section~\ref{experiment:ctr}, we only include the first of the six
months of MerRec data. The experiment was conducted using the initial month of MerRec data, with a subset comprising $10\%$ of the total sequences to manage computational resources. This resulted in a dataset featuring 915,453 unique \verb|product_ids| and 885,678 distinct sequences. The experimental framework mirrored that of the CTR prediction task, with the last two interactions in a sequence earmarked for validation and testing, respectively.

\subsubsection{Baseline Setup.} Table \ref{tab:sbr_bench} shows the results of evaluating four widely cited SBR baselines: 
CNN-based NextItNet~\cite{nextitnet2019}, RNN-based GRU4Rec~\cite{gru4rec2016}, Transformer-based 
SASRec~\cite{sasrec2018} and Bert4Rec~\cite{bert4rec2019} on MerRec data. 
We evaluate these baselines with the standard top-$K$ (5 and 20 here) 
ranking metrics such as
normalized discounted cumulative gain (NDCG)~\cite{gru4rec2016} and Recall (hit ratio)~\cite{bert4rec2019}.
Table~\ref{tab:hyperparamter_sbr} shows the hyperparameter setup for these experiments. We utilize an Nvidia L4 GPU 
instance with 16 cores and 128 GB RAM running on Google Cloud Platform to run the training and 
evaluations.

\subsubsection{Implications.}
In the test metrics, we see that the bidirectional Bert4Rec under performs other unidirectional variations NextItNet, GRU4Rec, and SASRec
similar to the observed results by Yuan et al.~\cite{tenrec2022}. 
We also see that GRU4Rec out performs SASRec for all the metrics, this is 
different than the observations in ~\cite{tenrec2022} underscoring the need
for evaluations with multiple benchmarks to understand the generality of 
model performances in varied settings.

\subsubsection{Challenges in C2C Marketplaces.}
\label{experiment:sbr-challenges}
All the above mentioned SBR models are variations of traditional language models where just like a language
model, at each time step, instead of predicting a token (corresponding to a word), these models
predict a token (corresponding to an \verb|item_id|). This modeling makes an implicit assumption that all
the tokens (\textit{i.e.} words or items) remain consistent during training, evaluation and serving. Any new token
is assigned an implicit <unk> characterization or broken into smaller known characters (\textit{e.g.}, Byte-Pair Encoding~\cite{bpe2016})
to assign an embedding to it. Note that the model will not be able to predict this new token
only process it as an input, since this token isn't burned into the model output space yet. 
This assumption is valid only for natural languages where new words are added less 
frequently, or platforms where items are static and can be consumed multiple times, like
various B2C e-commerce platforms or streaming services where each item is a movie or a TV Show.

For Mercari's C2C marketplace, each item is a single quantity listing, where it can be purchased
only once and it needs to be removed from the predictions after being purchased. Additionally, with 
a large number of item listings created daily, the output space of the recommendation model will 
need to be updated constantly. For these reasons, existing language model based sequential recommenders
are ill-suited for a dynamic C2C marketplace like Mercari. 

\begin{figure}
  \centering
  \includegraphics[width=1.05\linewidth]{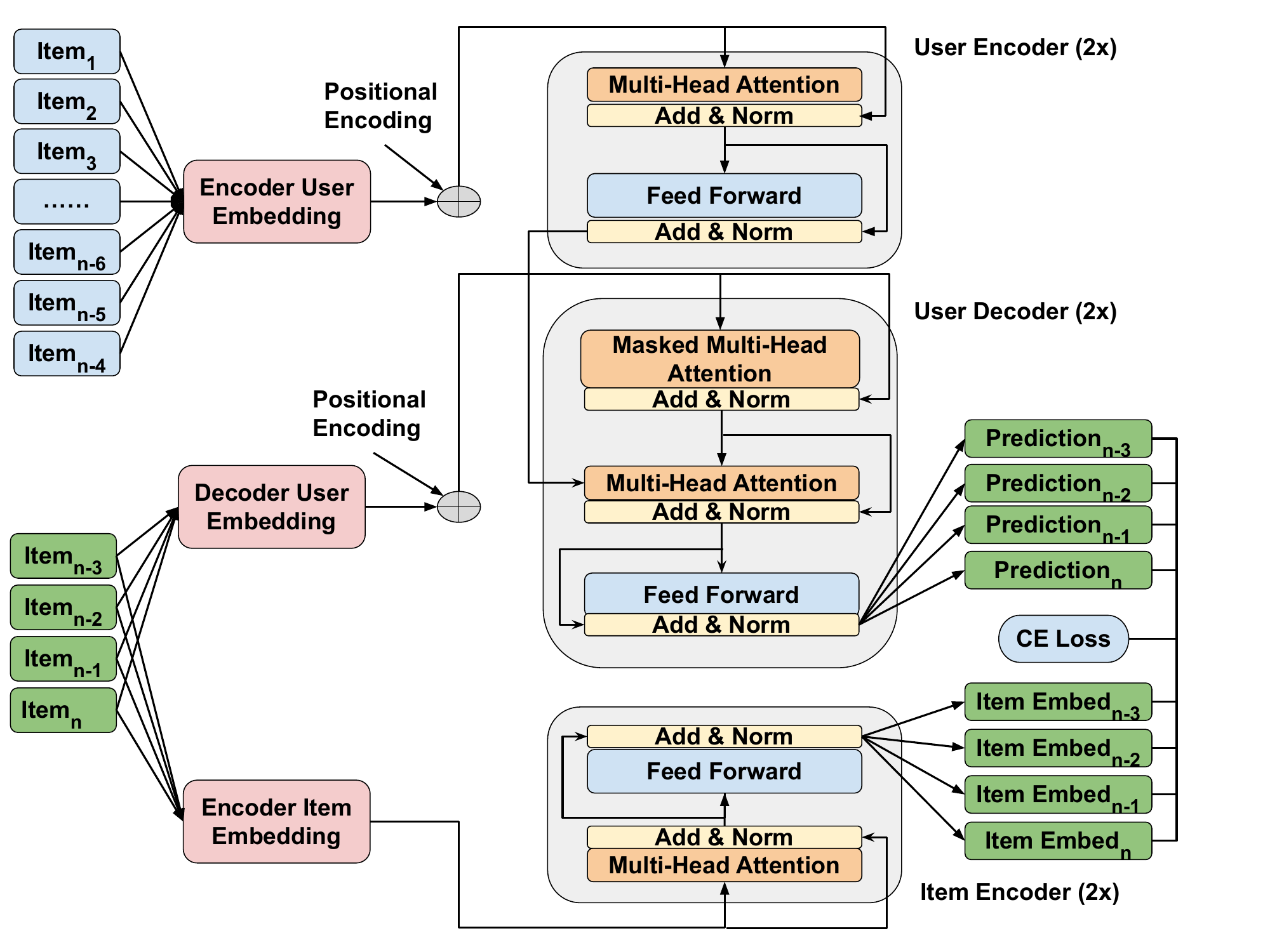}
  \caption{The Mercatran model employs a Three-Tower architecture with two user towers for predicting future interactions and one item tower for generating item embeddings. Each tower features two layers of specialized blocks to refine the embeddings, enabling the model to predict a user's next 4 interactions given up to 22 past interactions.}
  \label{fig:mercatran}
  \vspace{-0.5cm}
\end{figure}

\subsubsection{Mercatran.} 
\label{experiment:sbr-mercatran}
In response to the unique challenges of a Consumer-to-Consumer (C2C) marketplace, we developed Mercatran, a Three-Tower Transformer-based architecture. Inspired by the Two-Tower model used in YouTube recommendations by Covington et al.~\cite{covington2016}, Mercatran is tailored for dynamic marketplaces like Mercari to offer product-independent recommendations, leveraging user history to generate diverse and multistep recommendations.

Unlike the model proposed by Covington et al.~\cite{covington2016} where items (videos) are represented
by unique id's, in Mercatran, we encode items with the content features only, similar to Wang et al.~\cite{ebay2021}. The output of the model is modified to produce trained 
embeddings directly instead of tokens (corresponding to words in a language model). These trained embeddings can then be used to perform similarity search in a vector database to map the embeddings to items. 

Both Covington et al.~\cite{covington2016} and 
Wang et al.~\cite{ebay2021} use a two-tower architecture to encode users and items, however, we found
this architecture to be lacking in diversity where the recommendations generated are
similar to the last interacted item. The model is unable to learn from older interactions
even when using GRUs for the user tower~\cite{ebay2021}, since the output is a single embedding
which can't capture diverse user history. We built a solution conceptually similar to this called
Mercatran V1 as the precursor of our current implementation. While it was very successful
in driving orders, conversion and views (see Table~\ref{tab:sbr_online}), we observed
lack of diversity, where the model would only exploit the most recent interaction to recommend
items. 

To capture longer user-history and generate multi-time step recommendations, we propose a 
Three-Tower architecture called Mercatran V2, where we use Two-Towers to encodes users and a third tower to 
encode items. The two-towers to encode users are needed to generate recommendations
into the future auto-regressively, where the first tower is a standard transformer-based
encoder which can encode long user-history and the second tower is a transformer-based decoder
that can generate recommendations into the future. A third tower (a transfomer-based encoder, we call item encoder) is needed to provide
feedback to the user towers during training and it is fed the ground truth.
To train the model, we optimize a symmetric cross-entropy loss over the cosine
similarity scores of the predictions from the user-decoder and the associated predictions from the item-encoder, for each step, similar to Sohn~\cite{Sohn2016} and Radford et al.~\cite{clip2021}.

At inference time, the user towers are used to generate multi-step embeddings that can be used to query items in a vector database, while the third tower is used to create embeddings to index items in the vector database. The third tower can be a standard MLP. We use a transformer-based encoder without self-attention to function as the third tower. 

Figure~\ref{fig:mercatran} shows the overall architecture of our three-tower model.
In the current implementation, given user-history, at maximum 22 interactions, 
the model predicts the next 4 interactions (as embeddings), and these embeddings are used to perform a similarity search inside a vector database to find the recommended items. While this system does suffer from
sampling bias~\cite{jiyang2020, xinyang2019} due to the training data only containing
user interactions, generating multi-step recommendations introduces
diversity in the results aiding in providing more visibility to 
a more diverse set of items.

To increase the overall model efficiency in terms of the effective time horizon covered by events included in the attention window, we deduplicated consecutive events in the dataset (mentioned in Section \ref{data_description:cleaning}) if both the action type and item IDs remained the same per user, as we noticed that raw events without deduplication (e.g. buyers may repeatedly view the same item in consecutive events) often lead to a substantial reduction in sequential recommendation diversity and generally over-focusing on immediate-history trends. To further increase controllable diversity, we also force the model to generate multiple user encoding vectors when performing retrieval\footnote{During online tests, we also found engagement improvements by reusing cached snapshots of older user encoding vectors, as well as using a reranker that promotes more randomness in ranking}. This model-agnostic preprocessing is different from TransAct~\cite{xia2023transact} which tackled a similar problem at Pinterest but used a mix of model-side methods such as masking a random-length block of immediate-history events during training.

\begin{table}[]
\centering
\caption{Offline evaluation of Mercatran with 3 variations of input features, each measured at 3 levels of granularity, item (fine), category (coarse) and brand (coarse). The steps represent the 4 forecasted steps into the future. As can be seen
Mercatran maintains strong numbers even at 4 steps into the future.}
\label{tab:mercatran_offline}
\resizebox{\columnwidth}{!}{%
\begin{tabular}{c|c|ccccc|c|c}
\hline
\textbf{Input Features} &
  \textbf{Pred. Type} &
  \textbf{Step} &
  \textbf{nDCG @5} &
  \textbf{nDCG @20} &
  \textbf{Recall @5} &
  \textbf{Recall @20} &
  \textbf{Tr./Val Time (Hrs)} &
  \textbf{VRAM (GB)} \\ \hline
\multicolumn{1}{l|}{\multirow{12}{*}{Title + Brand + Category}} &
  \multirow{4}{*}{Item} &
  1 &
  0.0407 &
  0.0615 &
  0.0789 &
  0.1750 &
  \multirow{12}{*}{11.2} &
  \multirow{12}{*}{8.4} \\
\multicolumn{1}{l|}{} &                           & 2 & 0.0395 & 0.0598 & 0.0768 & 0.1702 &  &  \\
\multicolumn{1}{l|}{} &                           & 3 & 0.0380 & 0.0579 & 0.0743 & 0.1664 &  &  \\
\multicolumn{1}{l|}{} &                           & 4 & 0.0371 & 0.0567 & 0.0719 & 0.1627 &  &  \\ \cline{2-7}
\multicolumn{1}{l|}{} & \multirow{4}{*}{Category} & 1 & N/A    & N/A    & 0.4411 & 0.5746 &  &  \\
\multicolumn{1}{l|}{} &                           & 2 & N/A    & N/A    & 0.4427 & 0.5742 &  &  \\
\multicolumn{1}{l|}{} &                           & 3 & N/A    & N/A    & 0.4409 & 0.5740 &  &  \\
\multicolumn{1}{l|}{} &                           & 4 & N/A    & N/A    & 0.4398 & 0.5730 &  &  \\ \cline{2-7}
\multicolumn{1}{l|}{} & \multirow{4}{*}{Brand}    & 1 & N/A    & N/A    & 0.4952 & 0.6085 &  &  \\
\multicolumn{1}{l|}{} &                           & 2 & N/A    & N/A    & 0.4945 & 0.6072 &  &  \\
\multicolumn{1}{l|}{} &                           & 3 & N/A    & N/A    & 0.4934 & 0.6061 &  &  \\
\multicolumn{1}{l|}{} &                           & 4 & N/A    & N/A    & 0.4922 & 0.6052 &  &  \\ \hline
\multirow{12}{*}{Title} &
  \multirow{4}{*}{Item} &
  1 &
  0.0410 &
  0.0611 &
  0.0802 &
  0.1735 &
  \multirow{12}{*}{23.2} &
  \multirow{12}{*}{8.4} \\
                      &                           & 2 & 0.0402 & 0.0603 & 0.0780 & 0.1710 &  &  \\
                      &                           & 3 & 0.0402 & 0.0598 & 0.0786 & 0.1697 &  &  \\
                      &                           & 4 & 0.0399 & 0.0594 & 0.0771 & 0.1679 &  &  \\ \cline{2-7}
                      & \multirow{4}{*}{Category} & 1 & N/A    & N/A    & 0.4524 & 0.5988 &  &  \\
                      &                           & 2 & N/A    & N/A    & 0.4511 & 0.5981 &  &  \\
                      &                           & 3 & N/A    & N/A    & 0.4503 & 0.5974 &  &  \\
                      &                           & 4 & N/A    & N/A    & 0.4493 & 0.5975 &  &  \\ \cline{2-7}
                      & \multirow{4}{*}{Brand}    & 1 & N/A    & N/A    & 0.5024 & 0.6256 &  &  \\
                      &                           & 2 & N/A    & N/A    & 0.5031 & 0.6268 &  &  \\
                      &                           & 3 & N/A    & N/A    & 0.5038 & 0.6271 &  &  \\
                      &                           & 4 & N/A    & N/A    & 0.5037 & 0.6264 &  &  \\ \hline
\multirow{12}{*}{Brand + Category} &
  \multirow{4}{*}{Item} &
  1 &
  0.0111 &
  0.0196 &
  0.0217 &
  0.0609 &
  \multirow{12}{*}{23.2} &
  \multirow{12}{*}{8.4} \\
                      &                           & 2 & 0.0111 & 0.0194 & 0.0215 & 0.0593 &  &  \\
                      &                           & 3 & 0.0113 & 0.0193 & 0.0217 & 0.0583 &  &  \\
                      &                           & 4 & 0.0110 & 0.0189 & 0.0211 & 0.0575 &  &  \\ \cline{2-7}
                      & \multirow{4}{*}{Category} & 1 & N/A    & N/A    & 0.2769 & 0.3604 &  &  \\
                      &                           & 2 & N/A    & N/A    & 0.2810 & 0.3626 &  &  \\
                      &                           & 3 & N/A    & N/A    & 0.2818 & 0.3636 &  &  \\
                      &                           & 4 & N/A    & N/A    & 0.2806 & 0.3638 &  &  \\ \cline{2-7}
                      & \multirow{4}{*}{Brand}    & 1 & N/A    & N/A    & 0.3020 & 0.3485 &  &  \\
                      &                           & 2 & N/A    & N/A    & 0.3099 & 0.3587 &  &  \\
                      &                           & 3 & N/A    & N/A    & 0.3125 & 0.3643 &  &  \\
                      &                           & 4 & N/A    & N/A    & 0.3146 & 0.3680 &  &  \\ \hline
\end{tabular}%
}
\end{table}

\subsubsection{Serving Mercatran.} 
Mercatran is served using an open-source serving framework on a public cloud model prediction service. The prediction service can scale up to many nodes, depending upon traffic, where each node has 1 T4 GPU and 8 vCPUs and 30GB of RAM. With 1 such node, a single online inference to generate embedding for 1 item can take 40 - 50 MS on average, while a single inference to generate user embeddings for 1 user (note the model generates 4 embeddings) can take 60 - 70 MS on average. In general, many such nodes are ran online with load balancers, caching and fault tolerance handling to take on the enormous production traffic with minimal downtime.

To reduce the number of times Mercatran needs to be invoked at the time of request, we pre-compute user and item embeddings as much as possible with 2 offline asynchronous batch-jobs.
\begin{itemize}
    \item As users interact with our system, we pre-calculate embeddings corresponding to their interactions with Mercatran's User Towers, which are then stored in a feature store. Upon a user request, instead of invoking Mercatran directly, we first perform a lookup in the feature store to retrieve user embeddings.
    \item As new items are added into the system, Mercatran's Item Tower is used to embed these items in order to index them in a Vector Database. This is also performed in a batch setting
\end{itemize}
Note that these offline jobs are run close to real time with streaming in order to keep the recommendations fresh.

\subsubsection{Evaluation and Implication.}
To evaluate Mercatran offline, similar to the experimental setup in Section~\ref{experiment:sbr}, we train the model on the first of the six months of 
MerRec data. Unlike Section~\ref{experiment:sbr}, we train on a larger
number of sequences, since Mercatran doesn't model \verb|product_id|'s
explicitly but models items with content features, it needs more data to learn
meaningful representations. We filter out sequences smaller than 10 and only
use the last 22 events from larger sequences. Instead of just predicting the 
last interaction, as done previously, we train Mercatran to predict the last
4 interactions.
Ultimately we train on a sample of $8,042,410$ sequences from the 
first of the six months
of MerRec data, we evaluate on a sample of $403,821$ sequences from the second month
of data and test on a sample of $478,590$ sequences from the third month of data.
The training sequences consist of $27,231,493$ number of distinct items, 
while the evaluation consists of $1,532,388$ number of items and the 
test sequences consist of $1,653,843$ number of items.
For evaluation and testing, we index the last 4 interactions in the sequences
as items in a vector database. We use the remaining interactions to
autoregressively generate 4 vectors, which are then used to perform 
vector search in the created vector database. We measure the recall 
at 3 levels of granularity, i.e. category (coarse),
brand (coarse), and item (fine). We measure these numbers at 4 forecasted steps
labeled (1 to 4) to simulate predicting 4 future user-interactions.
Furthermore we measure nDCG @5 and nDCG @20 for item level predictions (fine)
at each of these 4 steps.
Table~\ref{tab:mercatran_offline} shows the full results of this evaluation. Please see Table~\ref{tab:hyperparamter_mercatran} for more information on the hyperparameters used in training Mercatran.

For ablation study on input features, we experimented on Mercatran with combination subsets of item title, brand and category label text. We see models trained on item title and item title + brand + category out perform models trained on only brand + category. This highlights the importance of item level features over just high level brand and category identifiers, since users might be interested in the some small subset within a brand or category not fully represented or discriminated by the identifiers alone. We do see that a model trained with only item title outperforms the model trained on item title, brand and category in many metrics. Since the difference is small, we think this can be attributed to the stochasticity introduced by sampling (since it’s not possible to evaluate on the entire marketplace) and/or feature redundancy. Despite this our production model uses all three due to additional information provided by the brand and category features.

Finally, we have deployed and tested two derivative versions of our Mercatran model on our production system online. Mercatran V1 (two-tower 1-step) was tested in October of 2022 against a purely brand plus category combination (i.e. the coarse proxy \verb|product_id|) based approach\footnote{The original $2022$ baseline was a short-sequence model that takes in both user demographic features and brief history of \texttt{product\_id}s from the user, and explicitly predicts the transitional probabilities among the various \texttt{product\_id}s. It then calls one of our search systems to retrieve final items.}, while Mercatran V2 (three-tower multi-step) was tested in April of 2023 against Mercatran V1. Both have seen great success in bringing positive changes to our user base. See Table~\ref{tab:sbr_online} for our performance differences from these two online tests.

\begin{table}[]
\caption{Two online production experiment results using our Mercatran models. V1 metrics are relative to Mercari's 2022 production baseline, and V2 metrics are relative to V1.}
\vspace{-0.2cm}
\label{tab:sbr_online}
\scalebox{0.85}{%
\begin{tabular}{@{}cccc@{}}
\toprule
\textbf{Model}          & \textbf{Orders} & \textbf{Conversion Rate} & \textbf{Views} \\ \midrule
MercatranV1 (2T 1-step) & +83\%           & +97\%                    & +12.1\%        \\
MercatranV2 (3T 4-step) & +124.4\%        & +23.8\%                  & +92.1\%        \\ \bottomrule
\end{tabular}%
}
\vspace{-0.5cm}
\end{table}

\vspace{-0.1cm}
\subsection{Multi-task Learning for Recommendation} \label{experiment:mtl}

\subsubsection{Task Description.}In multi-task learning we aim to learn multiple tasks simultaneously and attempt to maximize performance on a subset or all of them, sometimes striking balance or trade-off between tasks. This is practically interesting to a marketplace as users can perform many different actions to any item, therefore correctly anticipating the chance of each action can invite additional downstream user experience optimizations such as showing appealing alternative items, providing economical incentives, suggesting user to negotiate a counter offer or saving the item for consideration later depending on the business needs.

\subsubsection{Dataset Setup.}We also use the dataset compiled in the Section~\ref{experiment:ctr} but now it's applied to two tasks (anticipating \verb|item_view| and \verb|item_like| per item given user history and item contexts) instead of a single task. Note that MerRec provides 6 action types. That means our multi-task setting can be extended up to 6 tasks, according to the definition.

\subsubsection{Baseline Setup.}For example benchmark we chose two competitive classical models, for example OMoE~\cite{mmoe2018} and MMOE~\cite{mmoe2018} based on mixture of experts~\cite{moe1991} with additional gating mechanisms for all tasks, ESMM~\cite{ma2018entire} which models multiple tasks with explicit inter-task dependent relationships and a factorized loss function, as well as current state-of-the-art approaches such as AITM~\cite{xi2021aitm}, PLE~\cite{tang2020ple} and STEM~\cite{su2024stem}. In addition to the default two-task configuration, we also evaluated MMOE on single task (\verb|item_view|-only and \verb|item_like|-only) to show the difference against having two tasks simultaneously trained. Other methods such as ESMC~\cite{esmc2023}, RMTL~\cite{esmc2023}, PEPNet~\cite{pepnet2023} were omitted due to time and resource constraints. The ME-MMOE and ME-PLE competitive baselines introduced by the STEM~\cite{su2024stem} paper were omitted here as certain important implementation detail were not publicly available.

Similar to the CTR prediction (Section~\ref{experiment:ctr}) setup, here we also omit extensive hyperparameter search for time and resource reasons. See Table~\ref{tab:hyperparameter_mtl} and Appendix Section~\ref{hyper_alltasks} for hyperparameter tuning and configuration details. All executions in this subsection were ran on an Nvidia L4 GPU equipped linux system with 16 cores, 128GB RAM on Google Cloud Platform. See Table~\ref{tab:mtl} for the results using area under the curve (AUC) as the metric.

\subsubsection{Implications.}We observe that MMOE performs marginally better when using both tasks simultaneously than with either task in isolation, at the expense of overall lengthened training time. ESMM outperformed MMOE by insignificant amount on MerRec, whereas more pronounced performance gaps were observed on other benchmark datasets (e.g.~\cite{tenrec2022}). AITM, PLE, OMoE and STEM generally performed better and were easier to trigger early stopping during training, and there is a trend of performing better on the view task relative to the like task, whereas MMOE and ESMM performed better on the like task relative to the view task. STEM performed marginally worse than AITM, PLE and OMoE despite being the SoTA reported on other datasets~\cite{su2024stem}. This shows that MerRec has a highly diverse and challenging environment to tune and explore these types of models.

\begin{table}[]
\centering
\caption{Results for Multi-task Learning (MTL).}
\vspace{-0.2cm}
\label{tab:mtl}
\resizebox{\columnwidth}{!}{%
\begin{tabular}{cccccc}
\hline
\textbf{Model} & \textbf{view AUC} & \textbf{like AUC} & \textbf{Log Loss (View + Like)} & \textbf{Train+Val Time (Hrs)} & \textbf{VRAM (GB)} \\ \hline
Only view & 0.709 & N/A   & 0.395 & 16.3  & 3.93  \\
Only like & N/A   & 0.709 & 0.354 & 16.6  & 3.93  \\
MMOE      & 0.712 & 0.713 & 0.744 & 22.61 & 3.94  \\
ESMM      & 0.713 & 0.715 & 0.744 & 23.7  & 3.94  \\
AITM      & 0.773 & 0.736 & 0.684 & 2.85  & 3.26  \\
PLE       & 0.773 & 0.736 & 0.686 & 4.38  & 3.39  \\
OMoE      & 0.773 & 0.737 & 0.684 & 3.3   & 4.17  \\
STEM      & 0.772 & 0.736 & 0.683 & 9.68  & 11.27 \\ \hline
\end{tabular}%
}
\vspace{-0.85cm}
\end{table}

\subsection{Inference Acceleration for Recommendation}

\subsubsection{Task Description.}In real world use cases, recommendation predictions on marketplace can be pre-generated offline or live-generated online in real time. Regardless of the setup, we would like to study whether sequential recommendation models can enjoy inference speed ups to improve system efficiency\footnote{Experiments implemented here in a way that is comparable to past relevant academic recommendation literature, but inefficient compared to online production system standards. There are significant rooms for improvement in the program, and this preliminary study is only intended to examine potential relative speedup effects rather than serving as a basis to form generalized claims.}.

\subsubsection{Dataset Setup.}In this section we reuse exactly the same dataset format as Section~\ref{experiment:sbr}, down to the same seed, split ratio, etc.

\subsubsection{Baseline Setup.}Some work have shown that certain recommendation models can enjoy inference speedups while maintaining a similarly accurate performance. Chen et al.~\cite{skiprec2021} proposed the SkipRec framework which attaches a personalized policy network to learn a user-specific set of binary skip-or-keep decisions over the residual blocks of a modified NextItNet model backbone. They argue that by learning a user-specific block-selection policy, the overall computation is adapted to how ``easy'' or ``hard'' a user action sequence needs to be computed. SkipRec was introduced with two policy network design options based on the Gumbel-Softmax trick over all such binary decisions, or a self-critical sequence training (SCST)~\cite{scst2017} based reinforcement learning approach with curriculum learning~\cite{lookingforward2013, blockdrop2018} that incentivizes using minimal active blocks. Following the demonstration by SkipRec~\cite{skiprec2021} and TenRec~\cite{tenrec2022}, we study its effect on NextItRec~\cite{nextitnet2019} and SASRec~\cite{sasrec2018} using the Gumbel-softmax variant of SkipRec. All hardware configuration along with hyperparameters are identical to those used for Section~\ref{experiment:sbr} (except prediction batch size is now changed to 1 instead of 32, a practice we followed from~\cite{tenrec2022}) and we show our experimental results at Table~\ref{tab:infacc}. Please note that the total inference test wall time calculated is affected by numerous factors, such as batch size, test set size, etc.

\begin{table}[t]
\caption{Results of Inference Acceleration (IAR).}
\vspace{-0.2cm}
\label{tab:infacc}
\resizebox{\columnwidth}{!}{%
\begin{tabular}{@{}ccccccc@{}}
\toprule
\textbf{Model} & \textbf{nDCG @5} & \textbf{nDCG @20} & \textbf{Recall @5} & \textbf{Recall @20} & \textbf{Test Inference Time (Hrs)} & \textbf{VRAM (GB)} \\ \midrule
NextItNet       & 0.240 & 0.279 & 0.326 & 0.461 & 22.45 & 16.89 \\
Skip-NextItNet  & 0.195 & 0.229 & 0.265 & 0.383 & 22.45 & 15.11 \\
SASRec          & 0.152 & 0.183 & 0.207 & 0.314 & 21.84 & 12.51 \\
Skip-SASRec     & 0.209 & 0.242 & 0.281 & 0.397 & 21.64 & 14.40 \\ \bottomrule
\end{tabular}%
}
\vspace{-0.5cm}
\end{table}

\subsubsection{Implications.}We observed that when experimenting on the MerRec dataset, the overall speedup was small or practically negligible, but SkipRec's impact on the backbone's performance can be inconsistent: On NextItNet it can be too aggressive in skipping blocks which caused a degradation in NDCG and recall, while on SASRec it shortened the prediction time while boosted the NDCG and recall by significant amounts, indicating that the benefit or trade-offs of attaching SkipRec onto a sequential recommendation model backbone may be case dependent under the highly diverse MerRec dataset.

\section{Conclusion}

In conclusion, sourced from the Mercari platform, we propose the large-scale \includegraphics[height=\baselineskip]{figures/logo.pdf} dataset for Consumer-to-Consumer (C2C) recommendation. Our exploration into the MerRec dataset has highlighted the evolving role of recommendation systems in e-commerce, especially within the  marketplace. The introduction of the Mercatran model, tailored to address the unique challenges of C2C environments, represents a significant advancement in recommendation system research. Through tasks like click-through rate prediction, session-based recommendation, multi-task learning for user action prediction and inference acceleration in recommendation models, we've showcased the potential of the MerRec dataset and Mercatran to enhance user experience by accommodating the dynamic nature of C2C interactions.

The MerRec dataset's comprehensive capture of user-item interactions, combined with the innovative approach of Mercatran, opens new avenues for personalized and effective recommendation strategies in C2C platforms. This study not only bridges the gap between academic research and practical application but also sets the stage for future innovations in e-commerce recommendation.

\begin{acks}

We would like to deeply thank Junhyung Kim, Hayato Ryuki, Ian Jaffe, Fritz Mikio Kuribayashi, Majed Takieddine, Jonathan Rhodes, Kellen Miller, Joseph Burns, Brendan Gannon, Matthew Parrett, Abhishek Vilas Munagekar, Bethany Weaver, Shotaro Kohama, Raywin Stewart, Gabriel Driver-Wilson, Mohammad-Mahdi Mozzami for their immense support during the inception and development of the production versions of the Mercatran model and its efficient supporting  infrastructure. We would like to also thank James Hicks, Yoolim Jin and Kai Chun Cheung for their generous feedback and discussions regarding production deployment optimizations and implementation design.

\end{acks}



\appendix

\clearpage
\section{Experiment Configuration} \label{appendix_expr}

Here we provide additional details about the experiments. Table \ref{tab:hyperparameter_ctr} shows the hyperparameter configuration of the models chosen for the CTR experiment in Section \ref{experiment:ctr}, and Table \ref{tab:hyperparameter_mtl} for the MTL experiment in Section \ref{experiment:mtl}.

\subsection{CTR, MTL, SBR and IAR Hyperparameter Tuning (Excluding Mercatran)} \label{hyper_alltasks}

In CTR, MTL, SBR and IAR experiments (excluding Mercatran), we use a narrower set of candidate values per hyperparameter than those reported by the cited references, by picking the commonly reported best values (e.g. \cite{tenrec2022, su2024stem, mmoe2018} as well as most of the individual model-specific papers) and run grid search on them. For example, in the Multi-task learning section we observed from \cite{tenrec2022, su2024stem, mmoe2018} that most of the models often have either 16 or 32 for their embedding dimension, so these two values would become part of the grid search. We then repeatedly collect values like these for hidden dimension (e.g. for MLP layers in certain models like MMOE~\cite{mmoe2018} and OMoE~\cite{mmoe2018}), and other factors such as learning rate (e.g. 1e-3, 5e-4, 1e-4) for the Adam~\cite{kingma2015adam} optimizer and batch size (e.g. 1024, 2048, 4096). Other model-specific hyperparameters (such as the attention factor for AFM~\cite{afm2017}) are set to their single suggested value to limit the total amount of computation used.

One detail to note, is that we treat the embedding dimension slightly special, because the standard practice we observed is to use the same embedding dimension for all models on the same experiment for fairness, hence we try to manually look at the metrics across the spectrum on which embedding dimension brings higher values on average, then choose the better one to base the analysis on. One of the observations we had was that the embedding size tends to have a bigger impact on metric performance and training efficiency than most of the other hyperparameters, and selecting the other hyperparameters to be mostly the same values across models did not cause any significant discrepancies or unbalancing to the evaluation, as the performance variations across different other configurations were usually <1\% apart once seed is fixed. Similar phenomenons were also observed in other CTR~\cite{tenrec2022}, MTL~\cite{tenrec2022, su2024stem, mmoe2018}, and SBR~\cite{tenrec2022} studies to an extent.

\subsection{Mercatran Hyperparameter Tuning} \label{hyper_mercatran}

For Mercatran the hyperparameters are selected based on a variety of factors such as recall and nDCG as reported in the main manuscript but also cost. Given that the model is used to embed user interactions as well as to-be-indexed items coming into the platform, the smaller the embedding size, the smaller the storage needed. Hence we performed a grid search over the embedding size (with the goal to reduce the size) while also measuring the impact on the metrics simultaneously. For embedding size, we started with 512 and ultimately stopped at 64 below which we experienced significant degradation in our primary metrics. We used the same strategy for other parameters that also affect the cost such as number of stacked transformer blocks, sequence length, etc.

For parameters not related to the cost such as learning rate, batch size etc. as mentioned above, we perform grid search on most commonly reported parameters.

\begin{table}[htpb]
\centering
\caption{Hyperparameter setup of the CTR experiment.}
\label{tab:hyperparameter_ctr}
\resizebox{\linewidth}{!}{%
\begin{tabular}{@{}ccccc@{}}
\toprule
\textbf{Model} & \textbf{Embedding Size} $d$ & \textbf{Hidden Unit} $f$ & \textbf{Batch Size} $B$ & \textbf{Learning Rate} $\eta$ \\ \midrule
AFM     & 32 & 128 & 4096 & 0.00005 \\
NFM     & 32 & 128 & 4096 & 0.00005 \\
DCN     & 32 & 128 & 4096 & 0.00005 \\
DCNv2   & 32 & 128 & 4096 & 0.00005 \\
DeepFM  & 32 & 128 & 4096 & 0.00005 \\
xDeepFM & 32 & 128 & 4096 & 0.00005 \\
W\&D    & 32 & 128 & 4096 & 0.00005 \\ \bottomrule
\end{tabular}%
}
\end{table}

\begin{table}[htpb]
\caption{Hyperparameter setup for SBR experiments.}
\label{tab:hyperparamter_sbr}
\resizebox{\linewidth}{!}{%
\begin{tabular}{cccccc}
\hline
\textbf{Model} & \textbf{Embedding Size $d$} & \textbf{Hidden Unit $f$} & \textbf{Num Blocks $N$} & \textbf{Batch Size $B$} & \textbf{Learning Rate $\eta$} \\ \hline
NextItNet & 128 & 128 & 8  & 32 & 0.0001 \\
Bert4Rec  & 128 & 128 & 16 & 32 & 0.0001 \\
GRU4Rec   & 64  & 64  & 8  & 32 & 0.0005 \\
SASRec    & 64  & 64  & 8  & 32 & 0.0001 \\ \hline
\end{tabular}%
}
\end{table}

\begin{table}[htpb]
\caption{Hyperparameter setup for Mercatran}
\label{tab:hyperparamter_mercatran}
\begin{tabular}{cc}
\hline
\textbf{Parmeter}      & \textbf{Value} \\ \hline
Embedding Size $d$     & 64             \\
Feed forward Dim $d_{ff}$ & 1024           \\
Num Heads $h$          & 8              \\
Num Blocks $N$         & 2              \\
Batch Size $B$         & 3072           \\
Learning Rate $\eta$   & 1.0            \\
LR Decay Step          & 25             \\
Decay Gamma            & 1.0            \\
Vocab Size Limit       & 32768          \\ \bottomrule
\end{tabular}%
\end{table}

\begin{table}[]
\centering
\caption{Hyperparameter setup of MTL experiment.}
\label{tab:hyperparameter_mtl}
\resizebox{\linewidth}{!}{%
\begin{tabular}{@{}ccccc@{}}
\toprule
\textbf{Model} & \textbf{Embedding Size} $d$ & \textbf{Hidden Unit} $f$ & \textbf{Batch Size} $B$ & \textbf{Learning Rate} $\eta$ \\ \midrule
Only view (MMOE) & 32 & 128 & 4096 & 0.0001 \\
Only like (MMOE) & 32 & 128 & 4096 & 0.0001 \\
MMOE             & 32 & 128 & 4096 & 0.0001 \\
EESM             & 32 & 128 & 4096 & 0.0001 \\ \bottomrule
\end{tabular}%
}
\end{table}

\section{Dataset Details} \label{appendix_data}

\begin{figure*}
  \centering
  \scalebox{0.6}{\includegraphics[width=\linewidth]{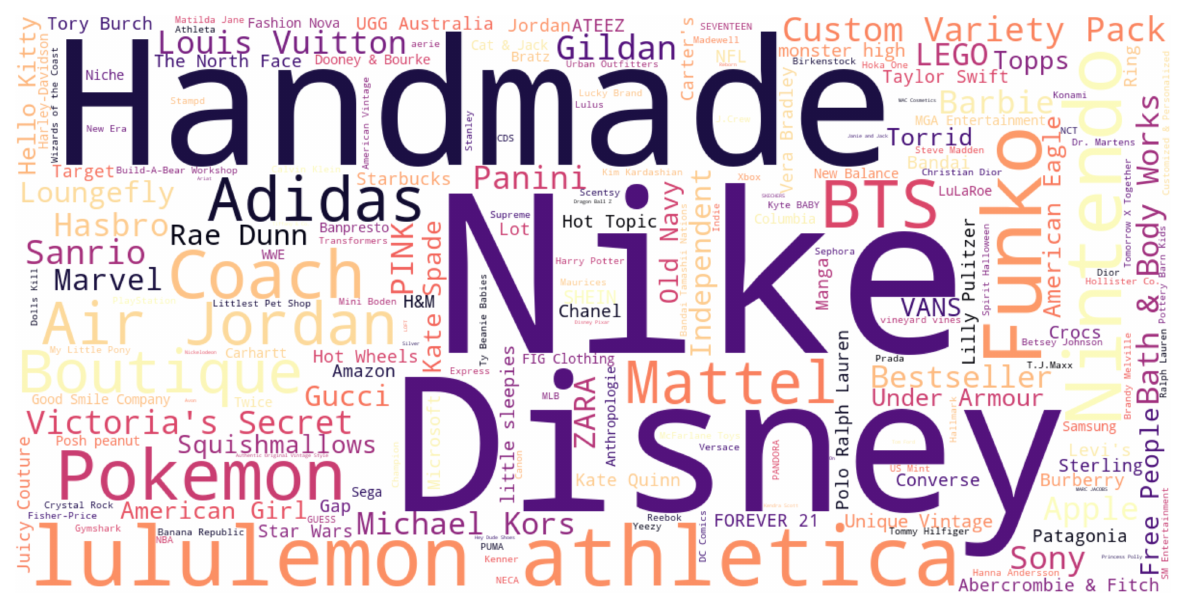}}
  \caption{Word cloud of the most frequently observed brands within MerRec. As implied here, MerRec (and in general, Mercari marketplace) has plenty of handmade (i.e. non-branded) items from sellers, which typically cannot reliably be associated with conventional product SKUs.}
  \label{fig:brand_wordcloud}
\end{figure*}

\begin{figure*}
  \centering
  \scalebox{0.45}{\includegraphics[width=\linewidth]{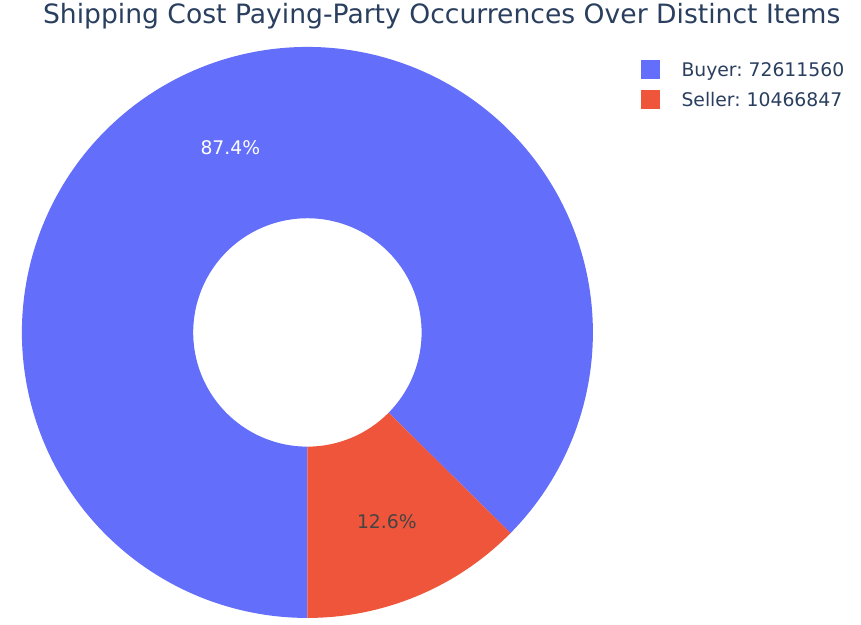}}
  \caption{Break down of shipment-paying party appearances over the distinct items.}
  \label{fig:shipper_counts}
\end{figure*}

\begin{figure*}
  \centering
  \scalebox{0.5}{\includegraphics[width=\linewidth]{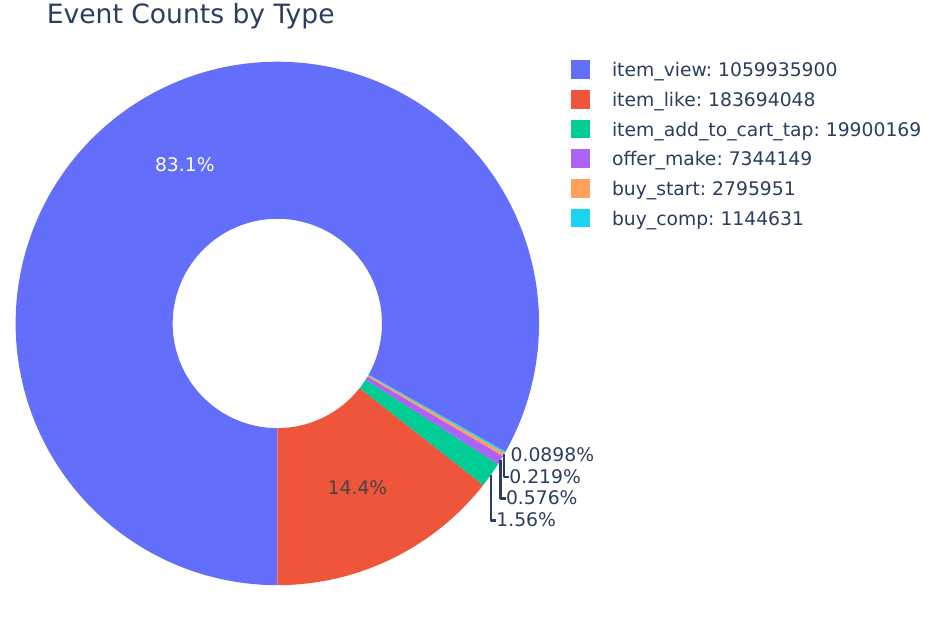}}
  \caption{Break down of the event instance counts by event ID. MerRec is primarily more focused on clicks largely due to nature of buyer exploration and browsing, but includes a few other action types.}
  \label{fig:event_counts}
\end{figure*}

\begin{figure*}
  \centering
  \scalebox{0.85}{
  \includegraphics[width=\linewidth]{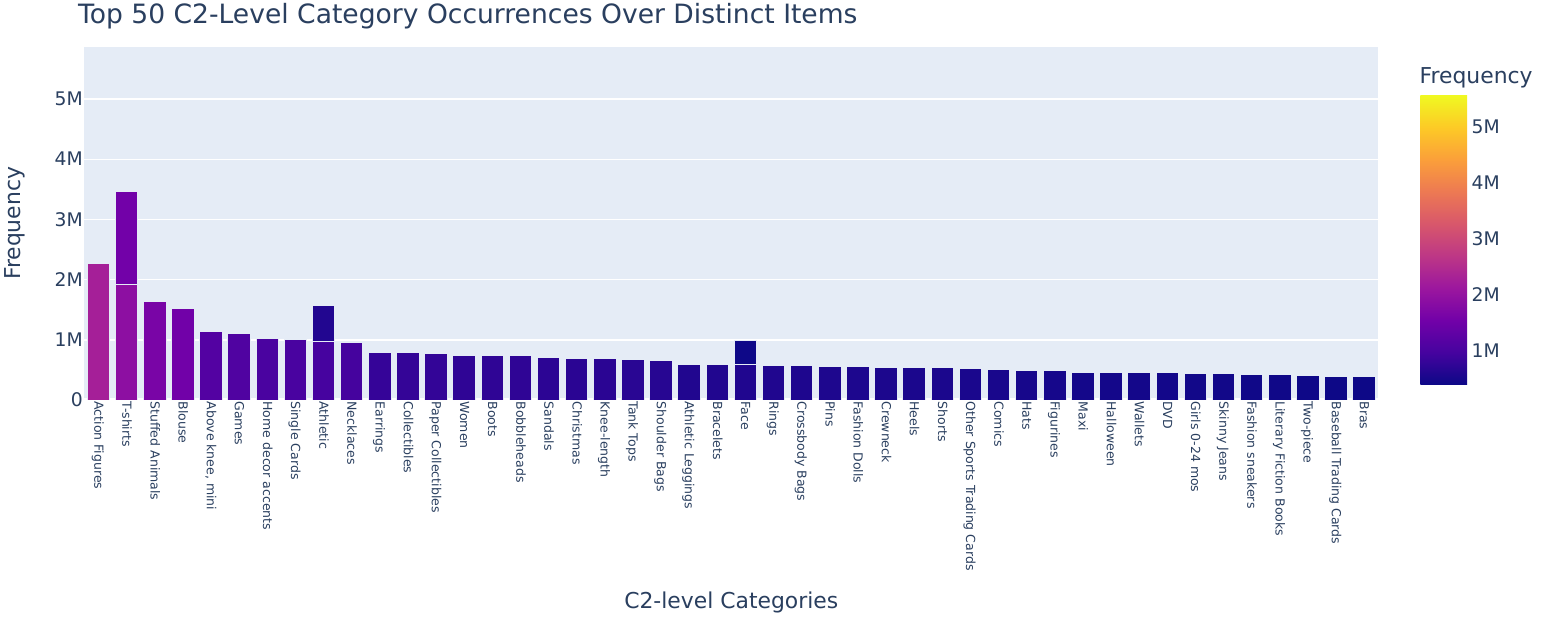}}
  \caption{Break down of the top 50 C2-level category appearances over the distinct items. The stacked bars represent categories which have the same name but originally belonging under different C0-level and C1-level categories.}
  \label{fig:c2_top50_counts}
\end{figure*}

\begin{figure*}
  \centering
  \scalebox{0.85}{
  \includegraphics[width=\linewidth]{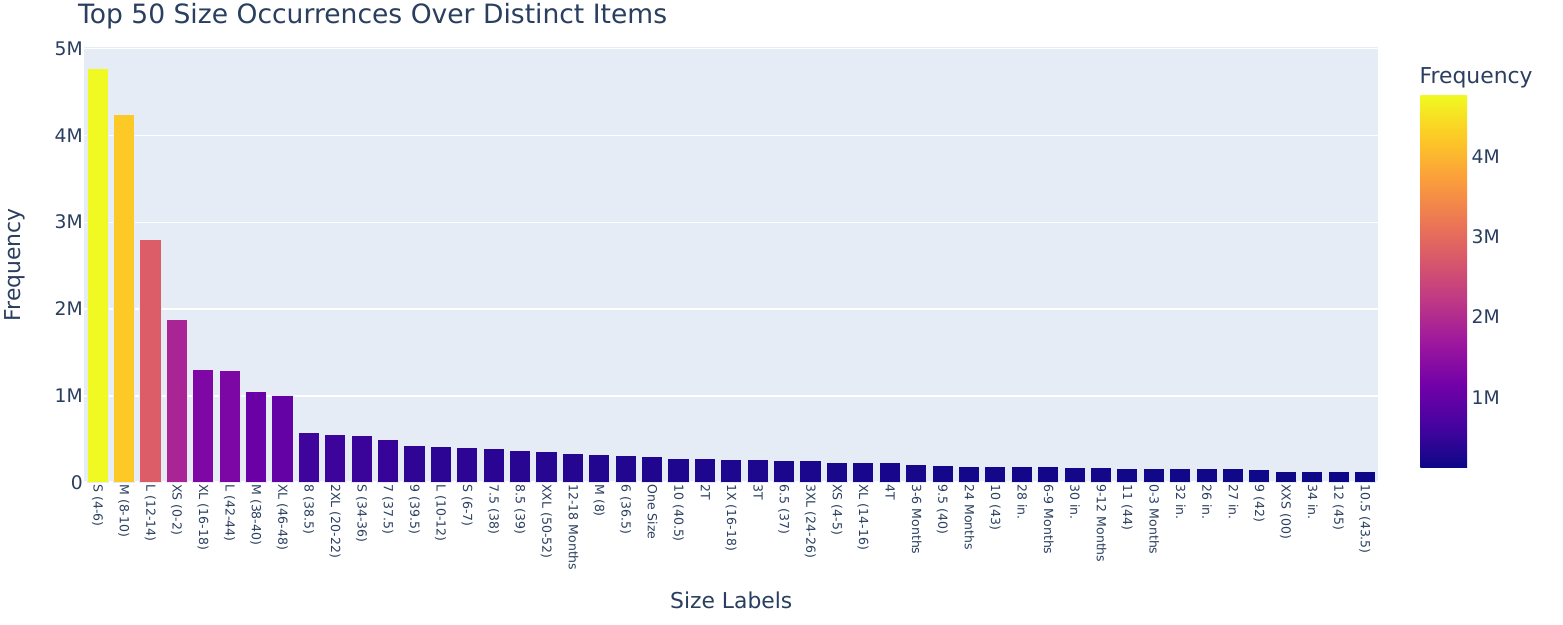}}
  \caption{Break down of the top 50 size label appearances over the distinct items. \textit{NULL}s are excluded here.}
  \label{fig:size_top50_counts}
\end{figure*}

\begin{figure*}
  \centering
  \scalebox{0.85}{
  \includegraphics[width=\linewidth]{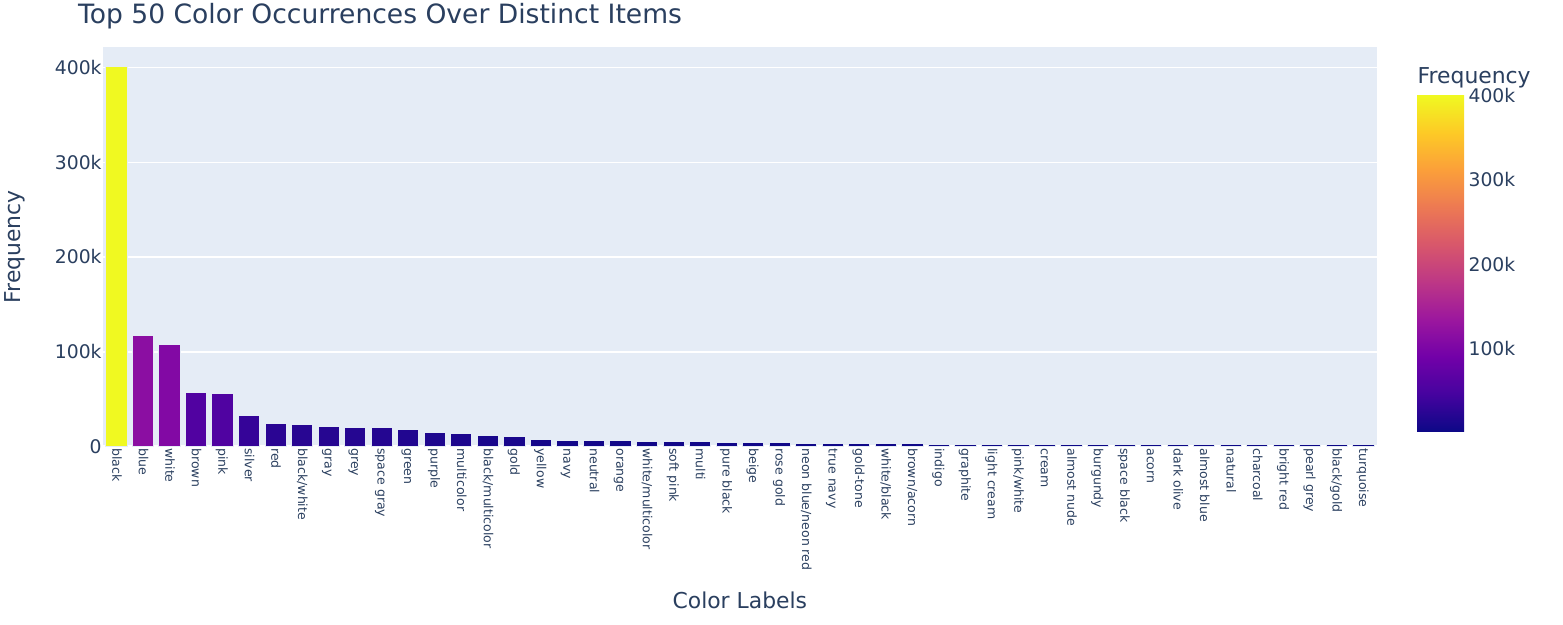}}
  \caption{Break down of the top 50 color label appearances over the distinct items. \textit{NULL}s are excluded here.}
  \label{fig:color_top50_counts}
\end{figure*}

In this section we provide additional details to help readers understand MerRec dataset better. Figure \ref{fig:c2_top50_counts} shows the most frequently observed C2 categories over distinct items in MerRec. Figure \ref{fig:size_top50_counts} displays the equivalent for size labels, and Figure \ref{fig:color_top50_counts} for color labels as well as Figure \ref{fig:shipper_counts} for which party covers shipping cost. Figure \ref{fig:brand_wordcloud} shows the word cloud of top observed brands among the distinct items. Figure \ref{fig:event_counts} reveals the breakdown of event counts by distinct event IDs.

\end{document}